\documentclass{aastex63}

\usepackage{graphicx}% Including figure files
\usepackage{amsmath}% Advanced maths commands 
\usepackage{amssymb}% Extra maths symbols                https://www.overleaf.com/project/5ecff7c0b894ba000149fb16
\usepackage{microtype}  % Makes minor kerning adjustments to make the layout beautiful to even the staunchest typesetting geek
\usepackage{url}

\newcommand{\degrees}{\ensuremath{^{\circ}}}

\received{\today}
\revised{this is submission V2}
\accepted{TBD}
\submitjournal{AJ}

\shorttitle{NICER Background Model}
\shortauthors{Remillard et al.}
\graphicspath{{./}{figures/}}
\begin{document}

\title{An Empirical Background Model for the \texttt{NICER} X-ray Timing Instrument}

\correspondingauthor{Ronald A.~Remillard}
\email{rr@space.mit.edu}

\author[0000-0003-4815-0481]{Ronald A. Remillard}
\affiliation{MIT Kavli Institute for Astrophysics \& Space Research, MIT, 70 Vassar St., Cambridge, MA 02139, USA}

\author{Michael Loewenstein} 
\affiliation{Department of Astronomy, University of Maryland, College Park, MD 20742}
\affiliation{X-Ray Astrophysics Laboratory, NASA Goddard Space Flight Center, Greenbelt, MD 20771, USA}

\author{James F. Steiner}
\affiliation{Smithsonian Astrophysical Observatory, 60 Garden St., Cambridge, MA 02138, USA}

\author{Gregory Y. Prigozhin}
\affiliation{MIT Kavli Institute for Astrophysics \& Space Research, MIT, 70 Vassar St., Cambridge, MA 02139, USA}

\author{Beverly LaMarr}
\affiliation{MIT Kavli Institute for Astrophysics \& Space Research, MIT, 70 Vassar St., Cambridge, MA 02139, USA}

\author[0000-0003-1244-3100]{Teruaki Enoto}
\affiliation{Extreme Natural Phenomena RIKEN Hakubi Research Team, Cluster for Pioneering Research, RIKEN, 2-1 Hirosawa, Wako, Saitama 351-0198, Japan}

\author{Keith C. Gendreau} 
\affiliation{X-Ray Astrophysics Laboratory, NASA Goddard Space Flight Center, Greenbelt, MD 20771, USA}

\author{Zaven Arzoumanian}
\affiliation{X-Ray Astrophysics Laboratory, NASA Goddard Space Flight Center, Greenbelt, MD 20771, USA}

\author{Craig Markwardt}
\affiliation{X-Ray Astrophysics Laboratory, NASA Goddard Space Flight Center, Greenbelt, MD 20771, USA}

\author[0000-0002-6575-2153]{Arkadip Basak}
\affiliation{Anton Pannekoek Institute, University of Amsterdam, Science Park 904, 1098XH Amsterdam, NL}

\author[0000-0002-5041-3079]{Abigail L. Stevens}
\affiliation{Department of Physics \& Astronomy, Michigan State University, 567 Wilson Road, East Lansing, MI 48824, USA}
\affiliation{Department of Astronomy, University of Michigan, 1085 South University Avenue, Ann Arbor, MI 48109, USA}
\altaffiliation{NSF Astronomy \& Astrophysics Postdoctoral Fellow}

\author[0000-0002-5297-5278]{Paul S. Ray}
\affiliation{Space Science Division, U.S. Naval Research Laboratory, Washington, DC 20375, USA}

\author[0000-0002-3422-0074]{Diego Altamirano}
\affiliation{Department of Physics and Astronomy, University of Southampton, Southampton, SO17 1BJ, UK}

\author{Douglas J.~K. Buisson}
\affiliation{Department of Physics and Astronomy, University of Southampton, Southampton, SO17 1BJ, UK}

\begin{abstract}

\texttt{NICER} has a comparatively low background rate, but it is highly variable, and its spectrum must be predicted using measurements unaffected by the science target. We describe an empirical, three-parameter model based on observations of seven pointing directions that are void of detectable sources. Two model parameters track different types of background events, while the third is used to predict a low-energy excess tied to observations conducted in sunlight.  An examination of 3556 good time intervals (GTIs), averaging 570 s, yields a median rate (0.4--12 keV; 50 detectors) of 0.87 c/s, but in 5\% (1\%) of cases, the rate exceeds 10 (300) c/s.  Model residuals persist at 20--30\% of the initial rate for the brightest GTIs, implying one or more missing model parameters. Filtering criteria are given to flag GTIs likely to have unsatisfactory background predictions. With such filtering, we estimate a detection limit, 1.20 c/s (3 $\sigma$, single GTI) at 0.4--12 keV, equivalent to $3.6 \times 10^{-12}$ erg cm$^{-2}$ s$^{-1}$ for a Crab-like spectrum. The corresponding limit for soft X-ray sources is 0.51 c/s at 0.3--2.0 keV, or $4.3 \times 10^{-13}$ erg cm$^{-2}$ s$^{-1}$ for a 100 eV blackbody. These limits would be four times lower if exploratory GTIs accumulate 10 ks. Faint-source filtering selects 85\% of the background GTIs, and higher rates are expected for targets scheduled more favorably. An application of the model to 1 s timescale makes it possible to distinguish source flares from possible surges in the background.

\end{abstract}



\section{Introduction} \label{sec:intro}

The \texttt{Neutron Star Interior Composition Explorer} (\texttt{NICER}) is a NASA mission for X-ray astronomy that has been operating on the \textit{International Space Station} (\textit{ISS}) since it was launched and deployed in 2017 June \citep{gend16}. The \texttt{NICER} X-ray Timing Instrument (XTI) consists of 56 identical and co-aligned cameras, each containing an X-ray Concentrator (XRC; \cite{okaj16}) and a customized Si drift detector positioned in the concentrator's focal plane.  The other primary components of \texttt{NICER} are a target acquisition and tracking platform and seven electronics boxes, each of which services event processing from eight detectors \citep{prig16}. Each detector package (detector, preamplifier, and thermoelectric cooler) is known as a Focal Plane Module (FPM), and each electronic box is referred to as a Measurement \& Power Unit (MPU).  The XTI sensitivity range is 0.2--12 keV, the energy resolution is typical of Si detectors (e.g., 150 eV FWHM at 6.5 keV), and detected events are time-tagged to an absolute accuracy of 100 ns. The combined detector output from the 50 best-performing FPMs offers substantial throughput, e.g. with 10,500 c/s from the Crab Nebula over the range 0.4--12 keV.

The FPM and MPU designs and interworking are described in \cite{prig16}.  Here we summarize the details that are most relevant to the background model at hand.  Each FPM is a single channel device that is collimated to view a circular celestial area with radius of 3.17 arcmin. The multilayer the collimator (1 mm radius) captures more than 90\% of the light in the concentrator's point spread function, and it limits the travel time to the anode for X-ray events, while the active area under the collimator extends to a radius of 2.8 mm. All \texttt{NICER} observations contain events from both the science target and the various types of background that are encountered while operating in space.  Scientific analyses thus require a model that can predict the background spectrum so that the target spectrum can be isolated. The 3C50 background model uses detector measurements that characterize the background but not the X-rays from the science target, analogous to past missions such as the Photon Counting Array of the {\it Rossi} X-ray Timing Explorer (RXTE) Mission \citep{jaho06}.   To assist modeling efforts, \texttt{NICER} routinely schedules observations of seven sky positions that are void of detectable point sources.  These targets (inherited from {\it RXTE}) are named ``BKGD\_RXTE\#'', with \# ranging 1--6,8.  Position \#7 was eliminated as a \texttt{NICER} background target for the presence of a soft X-ray source (bright star).

Pre-launch analyses predicted that the \texttt{NICER} background would primarily consist of a very small contribution from the cosmic diffuse X-ray background (e.g., \cite{wuha91}), given the small FOV of the Instrument (31.6 square arcmin), and particle interactions that deposit energy indistinguishable from in-band X-rays.  The 3C50 model covers these components.  Additional background sources that are not considered in the 3C50 model include enhanced diffuse X-rays from hot gas in the Milky Way (dependent on galactic latitude), possible soft X-rays related to Solar activity, and possible contamination from the Earth limb or the radiation sources on Soyuz spacecrafts.  In pre-launch analyses, the primary background components were expected to yield, for the majority of the \texttt{ISS} orbit, 0.2 counts per second (c/s) in soft X-rays at 0.4--2.0 keV, and an additional 0.15 c/s at 2--8 keV. 

The Empirical Background Model, also known as the ''3C50'' model, uses libraries constructed by sorting and combining the spectra extracted from background observations. Each library spectrum is the sum of spectra within a cell defined by intervals in the adopted model parameters, as described below. The model is named ``3C50'' because it is based on 3 parameters, the format assumes that spectral extractions will be made from standard \texttt{NICER} ``cleaned'' event lists, and the libraries are based on selection of 50 of the 52 FPMs operating in the XTI, while the remaining 4 (of 56) are not operating).

Our choice of model parameters (see below) requires additional introductory explanations about signal processing steps in the MPU (see \cite{prig16}).  The signal line for each FPM is replicated, so that events can be found and processed independently with ``fast'' and ``slow'' measuring chains that use circuits with different time windows.  The fast chain (84 ns nominal shaping time) produces time tags with higher precision, while the slow chain (465 ns shaping time) more effectively integrates the total electron yield, with lower noise, providing better measurements of the event energy.  Event detections can trigger on either measuring chain, when the rate of change in the signal line exceeds a trigger threshold held in the MPU, per FPM and per chain. The trigger thresholds are chosen to admit a noise rate of $\sim 3$ c/s per FPM, per measuring chain, as measured below 0.25 keV when there are no sources in front of the detectors. Such noise events will be asynchronous, i.e. they will trigger only one measuring chain.  On the other hand, X-ray and particle events will usually trigger both measuring chains, yielding a single event that includes two measurements of the event energy.  The caveat, here, is that the fast chain, with a higher noise level than the slow chain, has a lower trigger efficiency at energies below 1 keV. X-ray events from the source that trigger the slow chain, but not fast chain, are likely to be in the range 0.2--0.6 keV, i.e., above the slow threshold and below the fast threshold.

When the travel path in the detector is long, from the point of incidence to the charge-collecting anode at the center of the active Si region, then the size of the charge cloud, and hence the temporal profile of the event, is elongated by charge diffusion.  With its longer shaping time, the slow chain is more immune to such effects, since it has a longer time to integrate the charge. Incomplete charge collection will lower the reported event energy, and so the ratio of the slow chain energy and the fast chain energy systematically increases when the point of incidence is near the outer edges of the detector, i.e., beyond the inner ring of the collimator. This is an important detail for the background model, since pre-launch simulations had shown that particle interactions with the detector would generally produce event energies well above the 12 keV limit of the concentrator's effective area --- and thus be rejected --- except for edge-clipping events near the outer edges of the active Si area. \texttt{NICER's} calibrated events lists are given in the ''pulse invariant'' (PI) convention, where \texttt{PI} is the calibrated energy value from the slow chain in 10 eV units, \texttt{PI\_FAST} is the calibrated energy in the fast chain, and \texttt{PI\_ratio} = \texttt{PI} / \texttt{PI\_FAST}, for each event that triggers both of the measuring chains (and \texttt{PI\_ratio} = INDEF, otherwise). The events from the predicted edge-clipping particles that can mimic X-ray events in \texttt{PI} value must travel a long path to the anode, resulting in increased spread of the charge cloud associated with the increased drift time. The measurement of such events would then show anomalously high values in \texttt{PI\_ratio}, since the ``ballistic deficit'' in the fast chain will be more substantial, compared to the slow chain.

The final topic of \texttt{NICER} signal processing that is pertinent to the background model is the system of \texttt{NICER} event flags. There are six flags, with assigned value 1 or 0, which are interpreted as ``yes'' or ``no'', respectively.  Five of these flags tie an event to a particular circuit latch in the MPU, and the circuit is designed to help distinguish events as good or bad for inclusion in scientific analyses.  The flags are: first-event-in-packet (useful only for the data pipeline software), triggered the fast chain, triggered the slow chain, forced trigger, undershoot event, and overshoot event. Forced triggers result from commands to sample the signal values in the absence of a trigger from a detector, and they serve to monitor the zero point in the energy calibration of each FPM/MPU signal processing combination. Since launch, forced triggers have been operating at 5 Hz for each FPM.  Undershoot events have latched a circuit designed to detect the large negative pulse associated with a detector reset, which causes a high-amplitude negative pulse when an FPM discharges the capacitor that collects ambient charge running through the detector (maintained at $-55\degrees$C). Overshoot events have latched a different circuit designed to safeguard against large positive pulses (roughly equivalent to 18 keV) that could cause a bit rollover in the analog-to-digital converter. The keV assignments of undershoots and overshoot events are thus meaningless.  Any event with unity values in {forced triggers, undershoots, or overshoot} flags are excluded from the ``good events'' that are passed into the cleaned event lists created by the \texttt{NICER} pipeline.  The cleaned event lists are also limited to the slow chain energies in the range 0.2--15 keV.  Returning to the topic of background modeling, the strategy to recognize events associated with energetic particles can be summarized as follows.  Most particle events would be excluded as bad events via the overshoot flag, while detector edge-clipping events with in-band amplitude and no overshoot flag would be identified via high values in the event's \texttt{PI\_ratio}. As shown below,  \texttt{NICER} in-flight data reveal an additional component with neither of these properties that must be handled in order to predict the background spectrum.

\section{Observations and Data Selection} \label{sec:data}

The 3C50 model is a phenomenological approach to predict the in-band (0.4--12.0 keV) background spectrum using 
the observations of the \texttt{NICER} background fields. The parent data set for the model libraries includes all such observations from 2017 July 24, through 2020 March 21.  The contributions from each background field are given in Table~\ref{tab:obs}. The selection filters leading to the model libraries are summarized in Table~\ref{tab:select}, and they are described in detail, below.  

\begin{deluxetable*}{ccccc}
\tablenum{1}
\tablecaption{Parent Observations for the 3C50 Background Model \label{tab:obs}}
\tablewidth{0pt}
\tablehead{ \colhead{Target} & \colhead{ObsID first} & \colhead{ObsID last} & \colhead{\# GTIs} & \colhead{Exposure (ks)}}
\decimalcolnumbers
\startdata
BKGD\_RXTE1 & 1012010101 & 3012010106 & 357 & 131.2 \\
BKGD\_RXTE2 & 1012010201 & 3012020201 & 545 & 309.8 \\
BKGD\_RXTE3 & 1012010301 & 3012030104 & 317 & 157.2 \\
BKGD\_RXTE4 & 1012010401 & 2012040241 & 451 & 248.4 \\
BKGD\_RXTE5 & 1012010501 & 2012050232 & 540 & 292.4 \\
BKGD\_RXTE6 & 1012010601 & 3012060201 & 830 & 548.4 \\
BKGD\_RXTE8 & 1012010801 & 3012080102 & 516 & 336.9 \\
\enddata
\tablecomments{The sum of these background observations yields 3556 GTIs  and 2.024 Ms exposure time.} 
\end{deluxetable*}

Data analyses utilized the NASA HEASoft package, version 6.26.1. Prior software versions had different defaults for the use of \texttt{nimaketime} to define good time intervals (GTIs; see below). The raw event lists from the \texttt{NICER} pipeline for the observation IDs (ObsIDs) given in Table \ref{tab:obs} were calibrated for the 2020 gain revision specified by ``GCALFILE'', ``nixtiflightpi20170601v005.fits''.  When making the cleaned event lists, the default filters for environmental conditions were adopted, excluding times in the South Atlantic Anomaly (SAA), pointing elevations within 15\degrees\ of the dark Earth limb and within 30\degrees\ of the sunlit Earth limb. However, the default filters for bad-event count rates were effectively disabled by using extremely high values (15000) for the maximum rates of overshoots, undershoots, and the relationship between overshoots and magnetic cutoff rigidity.  These filter over-rides are required to populate the cleaned event lists with good events during times when the background rates are high.

\begin{deluxetable*}{lccl}
\tablenum{2}
\tablecaption{Data Selections and Filters for the Spectrum Libraries \label{tab:select}}
\tablewidth{0pt}
\tablehead{\colhead{Selection} & \colhead{\# GTIs} & \colhead{Exposure (ks)} & \colhead{Comment}}
\decimalcolnumbers
\startdata
All data                  &  3556  &  2024.2  &  from Table~1 \\
Selected 50 FPMs operating  &  3435  &  1937.4  &  \\
Filter noise outliers     &  3357  &  1891.1  &  any $i$: $nz_i > 100$ and $nz_i > 0.15 nz$ \\ 
Parameters within 3C50 limits &  3264  &  1818.3  &  see Fig.~\ref{fig:ibghrej} \\ \\
Subset \texttt{ISS} night  	      &  1991  &  1068.6  &  $nz < 200$, see Fig.~\ref{fig:cellrbg} \\
Filter out outliers           &  1947  &  1038.2  &  see Fig.~\ref{fig:cellrbg} ; GTIs for night library \\ \\
Subset \texttt{ISS} day            &  1273  &   758.6  &  $nz >= 200$ \\
Filter for high BG and stage 1 residuals  &  1076  &   627.4  &	 $ibg_{52} < 0.4$ c/s;  $-1.0 < (C_{net} + D_{net}) < 1.0$ \\
\enddata
\tablecomments{In summary, the 3C50 spectral libraries use 3023 (of 3556) GTIs, corresponding to 1665.6 ks or 82\% of the exposure time. There are 1947 GTIs contributing to the night library and 1076 GTIs for the day library.}
\end{deluxetable*}

The ObsID directories from the \texttt{NICER} pipeline contain data for a given target, accumulated on a given day. In this work, each GTI is an interval of continuous exposure, and for \texttt{NICER} such intervals are usually less than 2 ks because of interruptions imposed by the rotation of the \texttt{ISS} with respect to celestial coordinates, imposed by the rotation of the \texttt{ISS}, once per 93 min Earth orbit.  Many ObsIDs contain more than one GTI.  Since the \texttt{NICER} background can change significantly at different locations in the \texttt{ISS} orbit, background modeling is based on the timescale of GTIs, rather than ObsIDs.  There is further value to measuring the amount of parameter variability that occurs within a given GTI, so as to exclude it or to redefine the time boundaries to avoid strong flares in the background. Further practical considerations for running the background model are given in \ref{sec:practical}.

The tool \texttt{nimaketime} was used to define the GTIs for every ObsID in each of the background fields. This step repeats the same filter choices used to make the revised cleaned event lists (see above). GTI selection was additionally filtered to exclude GTIs with duration less than 60 s, while disregarding any gaps of 1 or 2 s that might be imposed by a telemetry packet loss, which is corrected for, via an adjusted exposure time, by the \texttt{NICER} pipeline.  The numbers of selected GTIs and the net exposure time accumulated per background field, are given in Table~\ref{tab:obs}. The total yield is 3556 GTIs, averaging 570 s per GTI and accumulating 2.024 Ms.

Finally, we choose to build the background spectrum libraries with a selection of 50 (of 56) FPMs. We label FPMs with two digits: the first for the MPU that services it (0-6) and the second for the FPM slot (0-7) in the MPU.  For example, the first FPM on the first MPU is ``00'', while the last FPM on th4e last MPU is ``67''. In this notation, the six excluded FPMs are the four that are not operating (11, 20, 22, and 60) plus two (14 and 34) that have shown episodes of unreliable spectra and high noise rates, respectively. The selection of the remaining 50 FPMs adds an element of uniformity to the background libraries.  However, users can conduct target analyses with any number of selected FPMs, and then apply the 3C50 model under the assumption that the model metrics do not change, detector-by-detector, and the net background spectrum can be simply scaled by the number of selected FPMs.

\section{First Stage of the 3C50 Background Model} \label{sec:ngt}

The strategy behind the 3C50 model is to sort out the background components from an empirical point of view, based on event properties found in observations of the background fields.  The first stage of the model distinguishes background components that have different spatial properties in the detector focal plane, as determined by event distributions that have different values of \texttt{PI\_ratio}.  The second stage of the model deals with the soft X-ray excess due to noise encroachment induced by the seepage of sunlight into the Instrument when observations are conducted in \texttt{ISS} sunlight, and this is described in Section~\ref{sec:day}. 

\subsection{Two Parameters for Background Components Sorted by  \it{PI\_ratio}}

\begin{figure}[ht!]
\includegraphics[angle=-90.,width=5.5in]{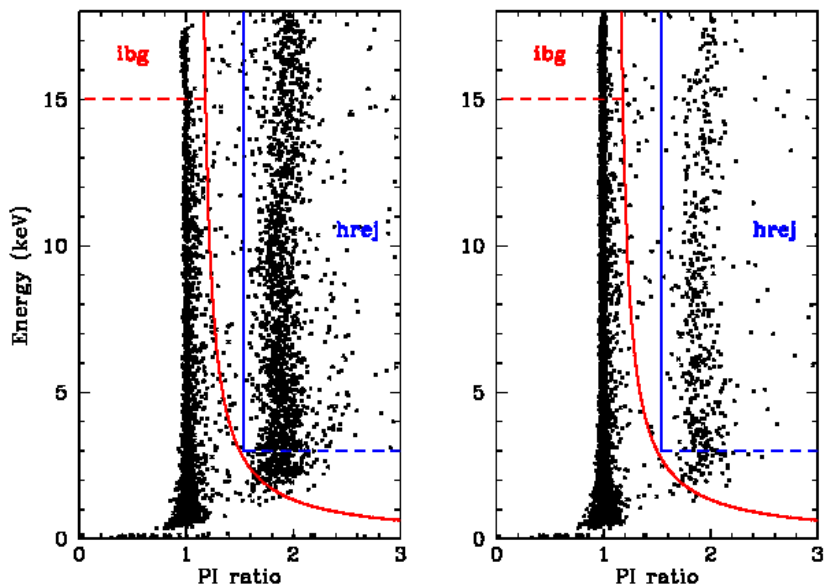}
\caption{Sample background observations showing every good event in the plane of event energy vs. \texttt{PI\_ratio}. The curved red line shows the \texttt{NICER} pipeline's boundary between events that appear as if in-focus (left of line), versus events originating far from the detector anode (right of line).  The integration areas for the first two background model parameters, $ibg$ and $hrej$, are also shown.  The observations were chosen to illustrate how the two background components are always present, but the fraction of events in each group can vary widely.
\label{fig:piratio}}
\end{figure}

The description of model parameter choices is framed by an examination of Fig.\ref{fig:piratio}, where events are plotted in the plane of (effective) photon energy (\texttt{PI}) versus \texttt{PI\_ratio}, i.e. the ratio of slow-chain to fast-chain keV values for good events that trigger both measuring chains. Each panel shows vertical spectral tracks centered on \texttt{PI\_ratio} values near 1.0 and in the range 1.7--2.2.  The curved red line shows the relationship, \texttt{PI\_ratio} $= 1.1 + 120.0 /$\texttt{PI}, where \texttt{PI} is the slow-chain photon energy in units of 10 eV. This relationship is the standard cut in pipeline data processing, where events to the left of the line, plus events with \texttt{PI\_ratio} = INDEF, are passed to the cleaned event lists. The value of 1.1 limits the exclusion of events to cases where the fast chain energy differs by more than 10\% from the slow chain energy. The second term adds allowances for the broadening of the \texttt{PI\_ratio} values due to statistical noise, especially in \texttt{PI\_FAST}. We note that the events displayed in Fig.\ref{fig:piratio} are drawn from the unfiltered but calibrated event lists (\texttt{*ufa.evt}) in the pipeline's \texttt{$ObsID$/xti/event\_cl/} directory, which contains all good events (i.e., events with no undershoot, overshoot, or forced trigger flags), with no \texttt{PI\_ratio} screening.  

The pipeline's \texttt{PI\_ratio} filter effectively separates the two vertical distributions in background events seen in Fig.\ref{fig:piratio}. The right-side distribution is consistent with the expected particle hits near the edges of the Si drift detector that would mimic good events with anomalously high values in \texttt{PI\_ratio}.  On the other hand, the left-side events are indistinguishable from ``in-focus'' X-rays emitted by \texttt{NICER} targets, and we refer to this distribution as the in-focus background component.  This label is intended as a comparative reference rather than a provable statement that the XRC is involved in the process of detecting these events.  Further considerations about the origin of these events are given below.

To capture and monitor the rate of events with high \texttt{PI\_ratio}, we define a ``hatchet'' rejection line, shown in Fig.~\ref{fig:piratio} as the blue vertical line at \texttt{PI\_ratio} = 1.54.  This leads to the choice of the first 3C50 model parameter, $hrej$, which is the count rate of hatchet-rejected events, \texttt{PI\_ratio} $ >= 1.54$ in the range 3--18 keV. The high-energy cutoff represents an approximate maximum energy in the \texttt{NICER} calibration, while the lower limit (3 keV) avoids any overlap between in-focus and hatchet-rejected distributions.  The tail-off of rejected events below 3 keV represents the lower efficiency of the fast chain triggers at lower energy, exacerbated by the charge-diffusion pulse broadening that further decreases the probability of detection.  There is also laboratory evidence that the gain drops near the detector edges, affecting both measuring chains. However, the energy content of particles is sometimes sufficient to endure all of these effects and trigger a pulse with a telltale high value in \texttt{PI\_ratio}.  In this sense, the \texttt{NICER} detector /  electronics package has a built-in particle monitor, albeit with low efficiency.

\begin{figure}[ht!]
\includegraphics[width=4in]{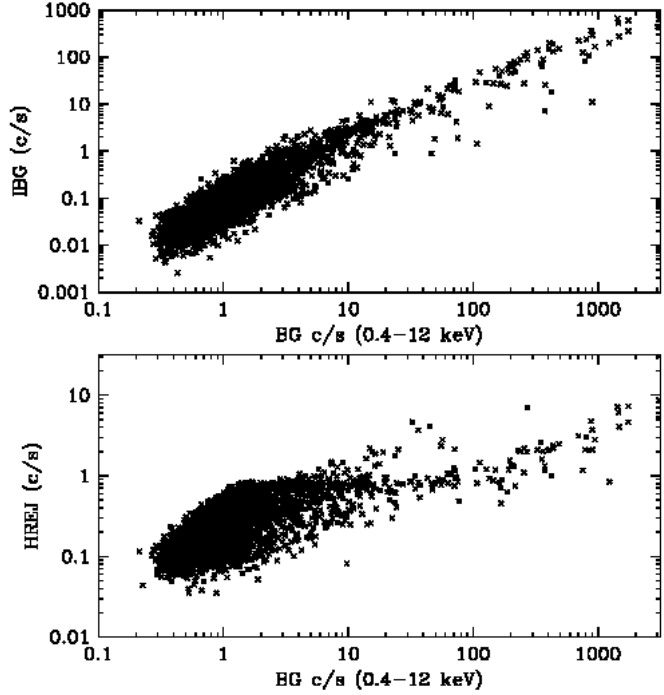}
\caption{The in-focus background rate at 0.4--12 keV ($R_{BG}$) versus the first two parameters of the 3C50 model, $ibg$ (top panel) and $hrej$ (bottom panel). The plotted data corresponds to all of the 3556 GTIs represented in Table \ref{tab:obs}. Rough correlations are seen in each panel, with a steeper dependence in the case of $ibg$. 
\label{fig:ampl}}
\end{figure}

Fig.~\ref{fig:piratio} shows two panels of background events, and each one represents an overlay from three GTIs, using different FPMs, selected at widely different times.  The GTIs for the left panel were chosen for having roughly half of the events located in the hatchet-rejected region, while the three GTIs selected for the right panel have $\sim 10\%$ of events in the hatchet-rejected region. The intent is to illustrate common features in the energy:\texttt{PI\_ratio} plane, while also showing that the fraction of events in each distribution can vary significantly. To monitor the rate of in-focus events from the background observations (i.e., the vertical track of events near \texttt{PI\_ratio} $\sim 1$ in Fig.~\ref{fig:piratio}), we define $ibg$, as the count rate of in-focus events (i.e., left of the red curve) in the range of 15--18 keV.  This restriction to high energy events is needed to limit the $ibg$ capture range to energies well above 12 keV, where the XRC optics have negligible effective area, to avoid contamination of $ibg$ by X-rays from bright sources.

Having defined $ibg$ and $hrej$ as the first two parameters for the 3C50 background model, Fig.~\ref{fig:ampl} shows how each parameter varies with the in-focus, in-band background count rate at 0.4--12 keV, hereafter $R_{BG}$. The lower limit of the energy range, i.e., choosing 0.4 keV instead of the XTI sensitivity limit at 0.2 keV, is a hedge against effects due to noise and the optical light leak, which are described in the next section. The primary objective of the background model is to predict the spectrum associated with $R_{BG}$.  Fig.~\ref{fig:ampl} shows that both $ibg$ (top panel) and $hrej$ (bottom panel) are roughly correlated with $R_{BG}$, with a steeper dependence in the case of $ibg$.  

The high degree of variability in $R_{BG}$ is apparent along the horizontal axes. The median value is $R_{BG} = 0.87$ c/s, but the distribution, even while ignoring the 1\% high and low extremes, still ranges from 0.33 to 300 c/s. Values of $ibg$ can also vary by many orders of magnitude. Since $ibg$ is the high-energy extension of $R_{BG}$), while the energy range is beyond the effective area of the \texttt{NICER} optics, $ibg$ values can be used to normalize the stage 1 library selection in the 3C50 model, tuning the model to converge with the source spectrum at 15--18 keV (considered in extrapolation, since extractions from cleaned event lists terminate at 15 keV under the default pipeline settings). 

While $hrej$ is a metric for the spatially extended events due to edge-clipping particles, the origin(s) of the in-focus background component, tracked with $ibg$, is not well understood.  Background components with \texttt{PI\_ratio} near unity are expected from the cosmic diffuse X-ray background, as well as possible soft X-ray emission from other sources (see Section \ref{sec:intro}). True X-ray events are expected to appear in-focus, since the metal collimator above the detector surface limits the path of X-rays to radii within 1 mm (3.17 arcmin) displacement from the anode \citep{prig16}. However, the count rate from these sources is expected to yield $R_{BG} \sim 0.5$ c/s over most of the sky, while measured $R_{BG}$ values are sometimes far brighter and highly variable. We are therefore led to view $ibg$ as representing a second particle component that is either unable to penetrate the collimator or is guided to the detector with assistance from the XRC. \\ \\

\subsection{3C50 Library for \texttt{ISS} Night}

\begin{figure}[ht!]
\includegraphics[width=4in]{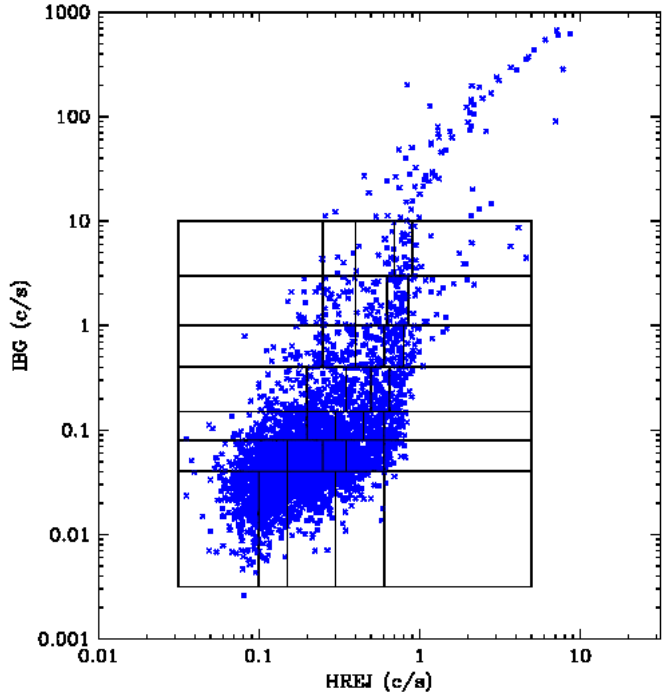}
\caption{Plot of $ibg$ versus $hrej$.  This plane is divided into five cells (horizontal) covering $hrej < 5.0$ and seven cells (vertical) with $ibg < 10.0$. The cells are referenced by ($ibg$, $hrej$) number with 11 on the lower left and 75 to the upper right.  Cell boundaries in units of $ibg$, $hrej$ are given in Table~\ref{tab:stage1}. These intervals  define bins in which the background spectra per GTI are combined to create the stage 1 background library for the 3C50 model. The stage 1 library contains 33 such spectra, with cells 15 and 71 unoccupied.
\label{fig:ibghrej}}
\end{figure}

Fig.~\ref{fig:ibghrej} shows a plot of $hrej$ versus $ibg$ for all of the 3556 GTIs represented in Table~\ref{tab:obs}. The $hrej$, $ibg$ plane is the parent for stage 1 of the 3C50 model.  A $5 \times 7$ grid of parameter values in this plane is used to bin the background spectra, per GTI, and the combined spectrum per grid cell is computed to populate the stage 1 library in the 3C50 model.  The cell boundaries are chosen to follow the population pattern, rather than a regular grid.  The library cells on the upper left and lower right of the grid are left vacant, and queries to those cells would select the nearest occupied neighbor, moving horizontally along $hrej$.

The stage 1 model cells are labeled with two digits: $ibg$ number (1-7) and $hrej$ number (1-5).  The layout is shown in Fig.~\ref{fig:ibghrej}. The cell boundaries are given in Table \ref{tab:stage1}, along with the average $R_{BG}$ values, the cell accumulation time, and the exposure-weighted normalization values in $ibg_{lib}$. Library selections must be standardized to a fixed number of FPMs, here chosen to be the maximum user choice of 52. However, users are free to select and specify, for any GTI$_i$, any number of FPMs ($nfpm_i$), along with the model parameter measurements, $ibg_i, hrej_i$.  The 3C50 model will then map a given GTI to a particular library cell, using values ($ibg_{52} = ibg_i * 52 / nfpm_i; hrej_{52} = hrej_i * 52 / nfpm_i$), and the matching library spectrum is presumed to have the correct spectral shape for that GTI. The matching library spectrum is then re-normalized by a factor $ibg_i / ibg_{lib}$, and the result is the stage 1 prediction for the background spectrum in the 3C50 model.

\begin{deluxetable*}{crrccccc}
\tablenum{3}
\tablecaption{Stage 1 Cell Boundaries and Normalizations \label{tab:stage1}}
\tablewidth{0pt}
\tablehead{ \colhead{Cell} & \colhead{$R_{BG}$} & \colhead{GTI time} & \colhead{$ibg_{52}$ start} & \colhead{$ibg_{52}$ end} & \colhead{$hrej_{52}$ start} & \colhead{$hrej_{52}$ end} & \colhead{$ibg_{lib}$} norm. \\
\colhead{($ibg, hrej$)} & \colhead{(c/s)} & \colhead{(ks)} & \colhead{(c/s)} & \colhead{(c/s)} & \colhead{(c/s)} & \colhead{(c/s)} & \colhead{(c/s)} 
}
\decimalcolnumbers
\startdata
11 & 0.446 &  44.3 & 0.00 & 0.04 & 0.00 & 0.10 & 0.0218 \\
12 & 0.493 & 149.4 & 0.00 & 0.04 & 0.10 & 0.15 & 0.0265 \\
13 & 0.596 & 100.3 & 0.00 & 0.04 & 0.15 & 0.30 & 0.0285 \\
14 & 0.775 &  28.5 & 0.00 & 0.04 & 0.30 & 0.60 & 0.0317 \\
15 &  ...  &  ...  & 0.00 & 0.04 & 0.60 & 10.0 & ... \\
21 & 0.689 &  95.7 & 0.04 & 0.08 & 0.00 & 0.15 & 0.0503 \\
22 & 0.801 & 109.8 & 0.04 & 0.08 & 0.15 & 0.25 & 0.0517 \\
23 & 0.855 &  71.4 & 0.04 & 0.08 & 0.25 & 0.35 & 0.0527 \\
24 & 1.008 &  67.5 & 0.04 & 0.08 & 0.35 & 0.60 & 0.0567 \\
25 & 1.242 &   8.0 & 0.04 & 0.08 & 0.60 & 10.0 & 0.0583 \\
31 & 1.103 &  41.0 & 0.08 & 0.15 & 0.00 & 0.20 & 0.1006 \\
32 & 0.955 &  29.5 & 0.08 & 0.15 & 0.20 & 0.30 & 0.0987 \\
33 & 1.322 &  43.9 & 0.08 & 0.15 & 0.30 & 0.45 & 0.1013 \\
34 & 1.302 &  20.4 & 0.08 & 0.15 & 0.45 & 0.60 & 0.1012 \\
35 & 1.771 &  16.5 & 0.08 & 0.15 & 0.60 & 10.0 & 0.1088 \\
41 & 2.020 &  18.6 & 0.15 & 0.40 & 0.00 & 0.20 & 0.2047 \\
42 & 1.724 &  27.2 & 0.15 & 0.40 & 0.20 & 0.35 & 0.2370 \\
43 & 1.904 &  15.6 & 0.15 & 0.40 & 0.35 & 0.50 & 0.2128 \\
44 & 2.414 &  20.6 & 0.15 & 0.40 & 0.50 & 0.65 & 0.2314 \\
45 & 2.673 &  16.9 & 0.15 & 0.40 & 0.65 & 10.0 & 0.2318 \\
51 & 2.683 &   8.6 & 0.40 & 1.00 & 0.00 & 0.25 & 0.5002 \\
52 & 2.763 &  20.1 & 0.40 & 1.00 & 0.25 & 0.40 & 0.5659 \\
53 & 3.293 &  10.0 & 0.40 & 1.00 & 0.40 & 0.60 & 0.6039 \\
54 & 4.446 &   7.5 & 0.40 & 1.00 & 0.60 & 0.80 & 0.7110 \\
55 & 4.288 &   9.3 & 0.40 & 1.00 & 0.80 & 10.0 & 0.5986 \\
61 & 4.527 &   7.3 & 1.00 & 3.00 & 0.00 & 0.25 & 1.4975 \\
62 & 6.080 &  13.0 & 1.00 & 3.00 & 0.25 & 0.40 & 1.7515 \\
63 & 6.644 &   4.9 & 1.00 & 3.00 & 0.40 & 0.63 & 2.0238 \\
64 & 7.604 &  11.4 & 1.00 & 3.00 & 0.63 & 0.85 & 1.6836 \\
65 & 11.644 &  7.7 & 1.00 & 3.00 & 0.85 & 10.0 & 2.1318 \\
71 &  ...  &  ... & 3.00 & 10.00 & 0.00 & 0.25 &   ...  \\
72 &  9.944 &  2.5 & 3.00 & 10.00 & 0.25 & 0.40 & 3.8110 \\
73 & 16.477 &  3.4 & 3.00 & 10.00 & 0.40 & 0.70 & 5.2374 \\
74 & 15.311 &  2.9 & 3.00 & 10.00 & 0.70 & 0.90 & 3.6876 \\
75 & 17.046 &  4.4 & 3.00 & 10.00 & 0.90 & 10.0 & 4.3715 \\
\enddata
\end{deluxetable*}

\begin{figure}[ht!]
\includegraphics[angle=-90.,width=5.5in]{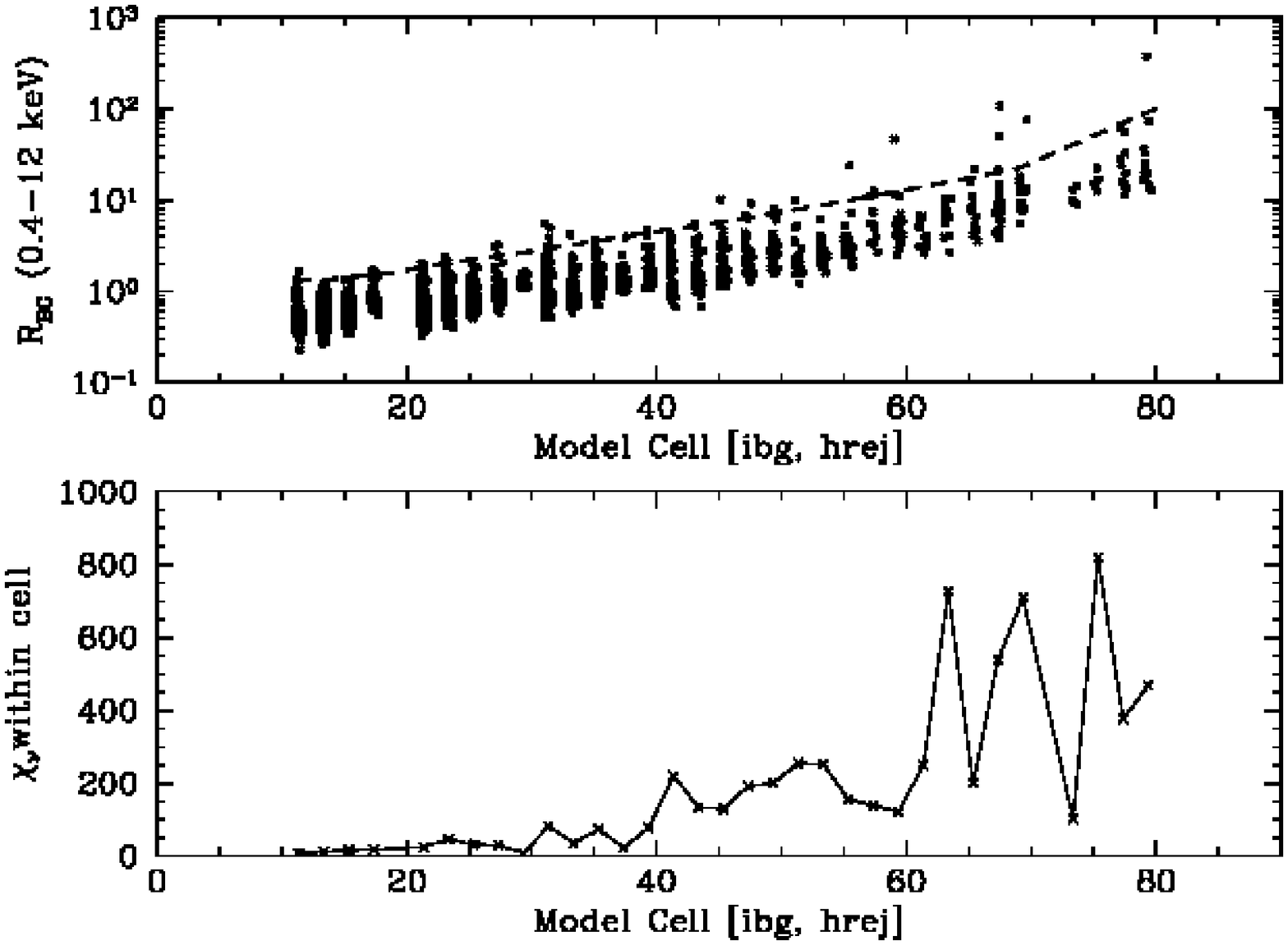}
\caption{top panel: Values of $R_{BG}$, the in-focus and in-band count rate for the background GTIs, are plotted for each successive cell of the stage 1 library.  The cells are counted by their $ibg, hrej$ number, ranging 11 to 75. The cell id numbers and their ordinal number within the cell are used to stretch out the points along the x-axis (see text). The dashed line shows the high-rate filter that clips the brightest outliers to be excluded when computing the library spectra. bottom panel: Reduced chi square values for each cell, after filtering out the high points (top panel), assuming that $R_{BG}$ is constant within each cell. The levels of intrinsic scatter in each cell are $20-40$ \%, motivating a step to normalize each library selection using the $ibg_i$ value of each target spectrum.
\label{fig:cellrbg}}
\end{figure}

Before proceeding to complete the stage 1 library spectra, we examine the distribution in $R_{BG}$ that is found in each cell, as seen in the upper panel of Fig.~\ref{fig:cellrbg}.  Only the 1991 GTIs sorted for \texttt{ISS}-night conditions (see Table~\ref{tab:obs}) are included here. Cells are noted by bin value, [$ibg: i=1-7, hrej: j=1-5$], and we use the cell number plus the fractional order of membership within a given cell to create an artificial horizontal axis that stretches out the data points simply for viewing purposes.  Data from cell i,j begin at value $10*i + 2*(j-1) + 1$, and the last GTI in the cell is plotted one unit later.  For example, the GTIa in cell 11 are plotted in the range $11 < x < 12$,
while cell 12 has range $13 < x < 14$, and cell 75 has range $79 < x < 80$.  The larger gaps indicate the vacant cells, 15 and 71. 

Even after filtering out the points above the dashed line in Fig.~\ref{fig:cellrbg}, the variations in $R_{BG}$ within each cell are much larger than the statistical uncertainties.  This is shown in the lower panel of Fig.~\ref{fig:cellrbg}, where the reduced chi square value ($\chi^2_\nu$) are shown for each cell, after the bright cases are removed, with the assumption that $R_{BG}$ is constant within a cell.  For cell 11, where the lowest $R_{BG}$ rates limit the statistical precision, we find $\chi_\nu = 8.7$, and $\chi^2_\nu$ increases for the cells with higher rates (i.e., better statistics).  The rms variations in $R_{BG}$ within each cell correspond to intrinsic fractional fluctuations in the range 20--40\% of the mean values.  This result suggests that the $ibg, hrej$ parameter scheme is far from a deterministic model, and additional background parameters are likely to be important. These results also motivate the strategy to normalize the library sections using the $ibg_i$ value of a given source spectrum, to mitigate against the fractional errors of the cell parents that would be otherwise inherited.  Thus, the 3C50 model assumes that the spectral shape per library spectrum is appropriate, while the normalization is fine-tuned to the target spectrum at hand.

\begin{figure}[ht!]
\includegraphics[width=5.5in]{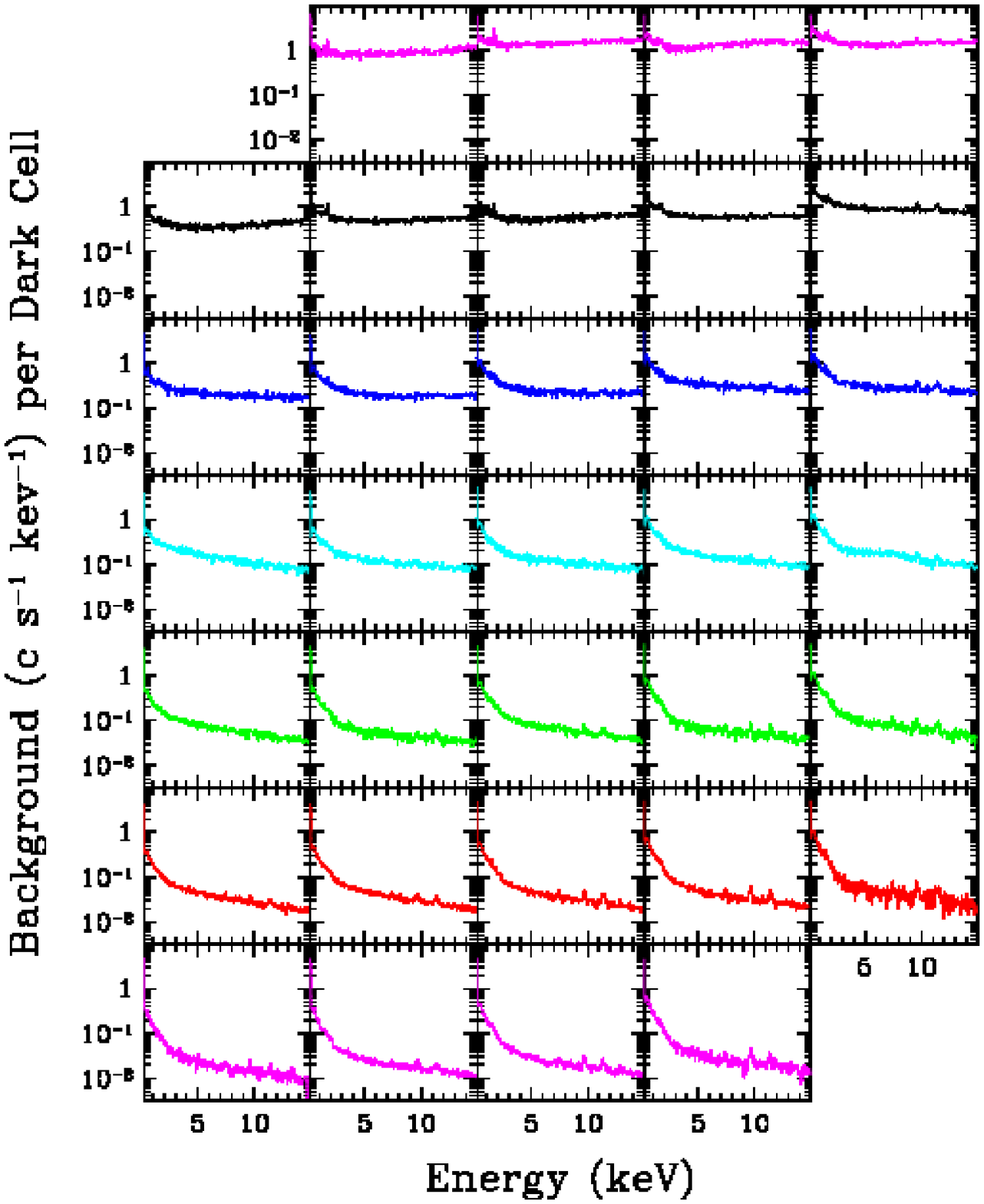}
\caption{The stage 1 library for the 3C50 background model consists of these 33 spectra arranged with $ibg$ number (1--7) increasing vertically and $hrej$ number (1--5) increasing to the right.  The spectra are substantially brighter and flatter with increasing $ibg$ cell number (vertical steps), while a more shallow brightness increase, a stronger soft component at 0.2--4 keV, and brighter emission lines are seen with increased $hrej$ cell number (left to right).  Cell 11 (lower left) is the closest that the library comes to an isolation of the cosmic diffuse X-ray background. 
\label{fig:lib1}}
\end{figure}

We note that for $R_{BG}$ or related quantities of central importance to background modeling, the exercise represented in Fig.~\ref{fig:cellrbg} can be conducted for any hypothetical set of model parameters. Intrinsic variances within cells and the progression of variances along each parameter axis can quantify how the cells organize the background measurements with minimal variance, and whether (from measurement slopes within each cell) the background properties are divided into a grid with sufficient resolution.

The library spectra are made for each cell after filtering out the data points that are above the dashed line in the top panel of Fig.~\ref{fig:cellrbg}, to avoid over-weighting the results by such cases. There is a library spectrum for each cell, which is simply the sum of all the counts in the selected GTIs of that cell, divided by the total exposure time. Table~\ref{tab:select} quantifies all of the selection and filtering steps used to construct the 33 spectra for the stage 1 library, using 1947 selected GTIs (of 1991 during \texttt{ISS} night) that are below the dashed line in Fig.~\ref{fig:cellrbg}.  Day and night assignments are made on the basis of the $nz$ value, as explained in the next Section. 

The stage 1 library spectra are displayed in Fig.~\ref{fig:lib1}, using a rebinning scheme that over-samples the XTI resolution uniformly by a factor $\sim 3$.  In the range 0.2--15 keV, the number of combined $PI$ bins is 3 (0.2--2.48 keV), 4 (2.48--6.00 keV), 5 (6.0--12.0 keV) and 6 (above 12 keV).  As expected, the spectra are substantially brighter and flatter with increasing $ibg$ cell number (vertical steps), and there is an appearance of the Si K--{$ \alpha$} emission line (1.74 keV) in the highest two levels of $ibg$. With increasing $hrej$ (i.e., from left to right), there is a more shallow increase in continuum brightness, a stronger soft component at 0.2--3 keV, and increasing emission lines indicating fluorescence at 7.47 keV (Ni K--{$\alpha$}), 9.71 keV (Au L--{$\alpha$}), and 11.44 keV (Au L--{$\beta$}). These changing spectral features over the surface of the $ibg, hrej$ plane offer some validation for the utility of choosing those model parameters. The background component tied to $ibg$ is the main source of the $R_{BG}$ count rate, while the $hrej$ parameter, despite its exclusion from $R_{BG}$ via $PI\_ratio$ filtering in the \texttt{NICER} pipeline, signals systematic changes in the spectrum of $R_{BG}$ for both the continuum shapes and the characteristics of emission lines. \\ \\

\section{Second Stage of the 3C50 Background Model} \label{sec:day}

The second stage of the 3C50 background model is required to subtract an independent soft X-ray component tied to observations during \texttt{ISS} daytime. The noise in the slow chain ($\sim 3$ c/s per FPM; see Section~\ref{sec:intro}) always creates a spectral component that is centered near 0.1 keV, and it is usually invisible at 0.3 keV.  However, it was recognized soon after launch that all of the \texttt{NICER} FPMs exhibit systematically higher levels of this low-energy noise when the XTI is illuminated by sunlight during the course of the \texttt{ISS} orbit (e.g, \citealt{bogd19}). Optical photons cannot trigger events in the MPUs, as such, but they liberate Si electrons, causing a number of secondary effects.  The increased detector current, in the presence of optical light, elevates the undershoot rate (i.e., detector reset rates) and also causes modest changes in detector gain and spectral resolution. The pipeline's gain calibration makes corrections for such effects, while the changes in spectral resolution are also predictable, again allowing appropriate corrections for science investigations. However, the spectral broadening of the low-energy electronic noise can increasingly intrude above 0.2 keV as the optical load becomes more intense. This is illustrated below in Subsection 4.2. Thus, the tail of the low-energy noise distribution can encroach on a portion of the in-band source spectrum during \texttt{ISS} daytime, making it necessary to include a quantification of this effect in the background model. There is no expected or measured correlation between excess noise and either $ibg$ or $hrej$, and so we treat the daytime soft excess as an independent spectral component to be handled in a second stage of the 3C50 model. When a GTI occurs during \texttt{ISS} daytime, the derived spectra from model stages 1 and 2 are simply added together to form the predicted background spectrum. \\

\subsection{Third Model Parameter for Soft X-ray Excess during \texttt{ISS} Daytime}

\begin{figure}[ht!]
\includegraphics[width=3.0in]{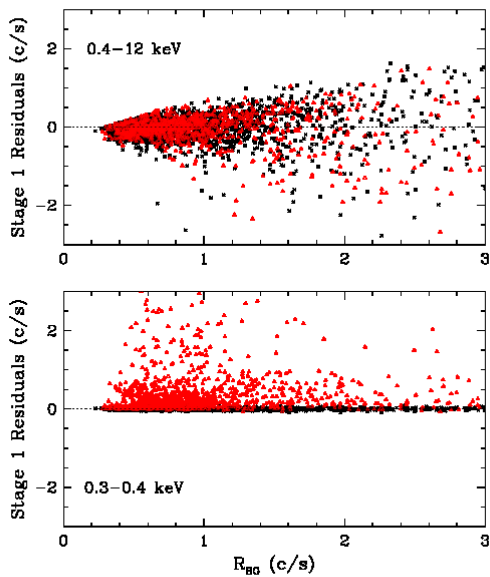}
\caption{The 3C50 model was applied to GTIs of the background observations using only the stage 1 library. Residuals from this exercise are shown at 0.4--12 keV (top panel) and 0.3-0.4 keV (bottom panel).  The plot symbols distinguish GTIs during \texttt{ISS} night (black cross) and \texttt{ISS} day (red triangle). The positive residuals during \texttt{ISS} daytime are associated with the seepage of sunlight into the XTI, which expands the $\sim 0.1$ keV noise component to the point that it can encroach into the soft X-ray region that is valuable for \texttt{NICER} science.
\label{fig:resid1}}
\end{figure}

The need to correct for low-energy noise in spectra obtained during \texttt{ISS} daytime is apparent in the residuals found after applying stage 1 (only) of the 3C50 model to the background spectra. Fig.~\ref{fig:resid1} shows the stage 1 residuals in two energy bands: 0.4--12 keV and 0.3-0.4 keV.  These model residuals should display an average value of zero, in all bands, if the background model has completed its job. GTIs during \texttt{ISS} night are plotted with a black cross, and GTIs during \texttt{ISS} day are plotted with a red triangle.  The stage 1 residuals are plotted versus $R_{BG}$, i.e., the original background rate at 0.4--12 keV. The horizontal axis is truncated at $R_{BG} = 3.0$, covering 89.8 \% of GTIs, for a better view of the details. Residuals in the range 0.4--12 keV (top panel) are fairly well contained during both day and night.  But in the range 0.3--0.4 keV (bottom panel), a region in soft X-rays that is valuable to \texttt{NICER} science investigations, the GTIs during \texttt{ISS} daytime show the encroachment of noise.  The vertical axis scale is chosen to be identical in both panels to highlight the significance of the problem.  

To incorporate the effect of the soft X-ray excess during \texttt{ISS} daytime, we choose to monitor the entire low-energy noise component for the 50 FPMs during each GTI. The third parameter for the 3C50 background model, $nz$ is defined as the total count rate in the slow chain in the range 0.0-0.25 keV.   The GTIs chosen for the stage 1 (nighttime) spectral library show a primary distribution peak in $nz$ at $156 \pm 5$ c/s, consistent with the $\sim 3$ c/s/chain target for trigger threshold settings for 50 FPMs (see \ref{sec:intro}). During the analyses leading to the 3C50 model, it was determined that the effects of optical light become measurable at 0.3--0.4 keV only when $nz > 200$ c/s. This value was then used to distinguish day and night categories for background GTIs in the 3C50 model. Quantification of the relationships between $nz$ and the soft excess in various energy bands are included in Table~\ref{tab:stage2}.

\subsection{3C50 Library for \texttt{ISS} Daytime}

An empirical strategy is adopted to model the soft excess during \texttt{ISS} daytime with an additional one-dimensional set of spectra that comprise the stage 2 library in the 3C50 model.  The stage 2 library captures the mean soft-excess spectra left behind by stage 1 of the background model, using 12 steps in value of $nz$, as given in Table~\ref{tab:stage2}.  This library strategy appears to offer better performance, compared to alternative efforts to fit the noise component with a function with broad wings, e.g., a Lorentzian or modified Gaussian. 

\begin{deluxetable*}{ccccccc}
\tablenum{4}
\tablecaption{Stage 2 levels and Quantified Soft Excess \label{tab:stage2}}
\tablewidth{0pt}
\tablehead{ \colhead{Level} & \colhead{Min. $nz_{52}$ c/s} & \colhead{Max. $nz_{52}$ c/s} & \colhead{Normalized $nz_{lib}$ c/s} & \colhead{S0 c/s} & \colhead{S1 c/s} & \colhead{A band c/s}}
\decimalcolnumbers
\startdata
01 & 200 & 215 &  202.81 &  0.159 &  0.011 & -0.010  \\ 
02 & 215 & 250 &  228.52 &  0.264 &  0.022 & -0.004  \\ 
03 & 250 & 300 &  277.36 &  0.491 &  0.037 &  0.001  \\ 
04 & 300 & 400 &  346.98 &  2.806 &  0.120 &  0.030  \\ 
05 & 400 & 500 &  457.51 &  2.091 &  0.141 &  0.007  \\ 
06 & 500 & 600 &  547.25 &  2.922 &  0.184 &  0.037  \\ 
07 & 600 & 750 &  683.16 &  5.032 &  0.307 &  0.019  \\ 
08 & 750 & 900 &  810.45 &  7.937 &  0.408 &  0.026  \\ 
09 & 900 & 1100 & 1003.81 & 12.466 &  0.632 &  0.061  \\ 
10 & 1100 & 1300 & 1198.35 & 31.528 &  1.230 &  0.114  \\ 
11 & 1300 & 1600 & 1393.30 & 60.267 &  1.901 &  0.132  \\ 
12 & 1600 & 0 & 1796.66 & 80.318 &  2.493 &  0.173  \\ 
\enddata
\tablecomments{The last 3 columns give the soft excess rates per 50 FPMs, and the energy bands are S0: 0.2--0.3 keV, S1: 0.3--0.4 keV, A band: 0.4--1.0 keV} 
\end{deluxetable*}

Starting with 1273 daytime GTIs, we apply two filters before combining the spectra within the designated levels in $nz$ (see Table~\ref{tab:stage2}).  Both filters are intended to reduce systematic problems that would be inherited by the stage 2 library.  The first filter limits the input GTIs to moderate count rates, using $ibg_{52} < 0.4$ c/s, which corresponds to the first four $ibg$ levels (of 7) in stage 1.  The second filter excludes cases in which the stage 1 residuals at 2--12 keV (i.e., away from the soft excess) are outside the range $\pm 1.0$ c/s.  This condition screens out the GTIs with stage 1 background spectra that deviate from the predicted one. These filters exclude 137 and 60 GTIs, respectively.  The parent spectra for the stage 2 library then consist of 1076 GTIs, amounting to 83\% of the total daytime exposure (see Table~\ref{tab:select}), divided into 12 $nz_{52}$ levels, as defined in Table~\ref{tab:stage2}.

\begin{figure}[ht!]
\includegraphics[width=4in]{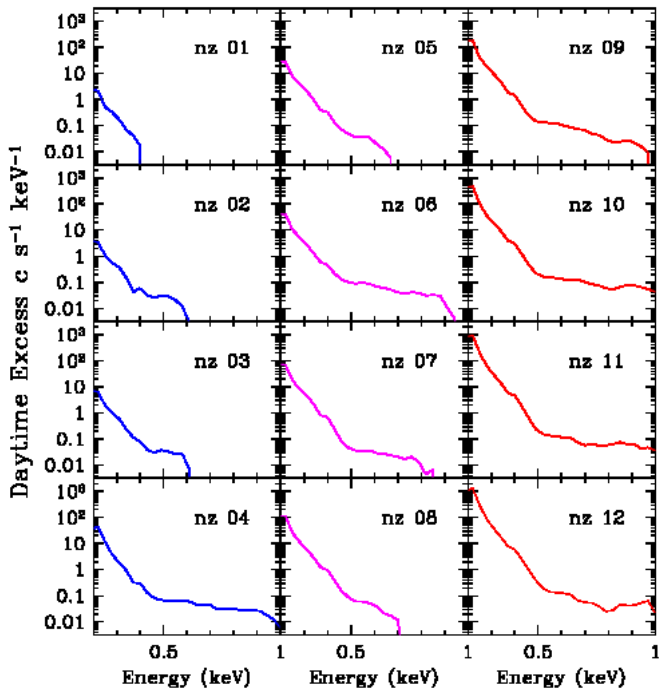}
\caption{The stage 2 library for the 3C50 background model consists of 12 spectra that represent the soft excess caused by various levels of sunlight intrusion into the XTI.  These spectra were determined from the residuals of the stage 1 model for 1076 GTIs during \texttt{ISS} day.  The 12 spectra correspond to different levels in the $nz_{52}$ parameter, which closely tracks the amount of the optical light leak.  
\label{fig:lib2}}
\end{figure}

The stage 2 model spectra are shown in Fig.~\ref{fig:lib2}. The amplitude and extent in keV of the soft excess increases with the $nz_{52}$ level, as expected.  These spectra have been smoothed with a 15 PI-bin ``boxcar'' at energies above 0.4 keV. Above 0.7 keV, the continuum levels are very faint, even at the highest $nz$ rates (i.e., levels 10--12), where the integrated count rates correspond to an addition of less than 0.06 c/s to the count rate (i.e., 0.7--12 keV) during \texttt{ISS} daytime. 

The normalization scheme for the stage 2 library selection follows the practices used for the stage 1 library.  For any spectrum ($GTI_i$), a user can select and specify any number of FPMs ($nfpm_i$), along with the noise level $nz_i$, that is consistent with $nfpm_i$. The stage 2 library intervals are again based on a standard 52-FPM scale (Table~\ref{tab:stage2}), so that a library selection is based on the value $nz_{52} = nz_i * 52 / nfpm_i$.  If $nz_{52}$ corresponds to level $j$ of the stage 2 library, then that spectrum is selected and re-normalized by a factor, $nz_i / nz_j$, and then added to the background prediction in stage 2 of the 3C50 model.

Table~\ref{tab:stage2} also specifies the count rates of library spectra integrated for the softest \texttt{NICER} bands, $S0$ (0.2--0.3 keV), $S1$ (0.3--0.4 keV), and $A$ (0.4--1.0 keV).  The Table quantifies the soft excess that would be suffered if stage 2 of the model were ignored.  Stage 2 is therefore a required part of the 3C50 model for studies of faint and soft X-ray sources, including the rotation-powered pulsars that are the prime targets for \texttt{NICER}, since their count rates are often $< 1$ c/s.  

Table~\ref{tab:stage2} can further help to evaluate residual count rates in background-subtracted spectra for science targets.
We retain our conventions for labeling \texttt{NICER} energy bands, and we use the subscript ``net'' to indicate count rate queries applied to background-subtracted spectra. The manner in which noise events leak into the different energy bands (Table~\ref{tab:stage2}) gives some guidance as to how to use $S0_{net}$ as a quality metric to estimate, per GTI, the likely level of residual contamination that may be present in $S1_{net}$, which we want to preserve for science.  This is relevant because systematic errors in the background model can leave behind many more counts in $S0_{net}$, compared to the X-ray brightness of the target, since the effective area of the \texttt{NICER} XTI is very low in $S0$ and this band is strongly attenuated by absorption in the interstellar medium.  Depending on the brightness and softness of a given X-ray target, such filtering steps using $S0_{net}$ can, for example, inform users as to whether the soft X-ray light curve (e.g., 0.3--2.0 keV) is free from contaminated GTIs, or when spectral fitting down to to 0.3 keV is likely to be safe. We offer specific recommendations and examples for filtering results via the background-subtracted spectra, in part using $S0_{net}$, in \ref{sec:practical} and \ref{sec:lc}, below.

\section{Model Evaluation} \label{sec:eval}

\begin{figure}[ht!]
\includegraphics[width=3in]{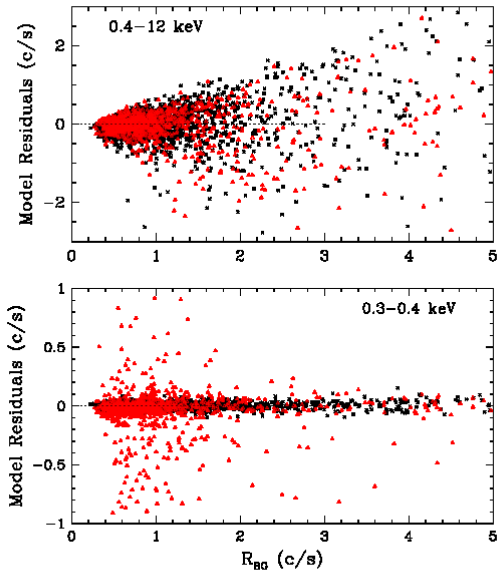}
\caption{Final residuals from the application of the 3C50 model to the background observations, plotted versus the original count rate at 0.4--12 keV, $R_{BG}$.  The horizontal axis is truncated at 5 c/s for clarity; only 6\% of the observations exceed this level. Model residuals are shown at 0.4--12 keV ($R_{BGnet}$, top panel) and 0.3-0.4 keV ($S1_{net}$, bottom panel).  The plot symbols distinguish GTIs during \texttt{ISS} night (black cross) and \texttt{ISS} day (red triangle).
\label{fig:resid2}}
\end{figure}

To close the loop on the background observations we apply the full 3C50 model to all of the 3477 background GTIs that have parameter values within the model limits. Residual count rates are shown vs. $R_{BG}$ in Fig.~\ref{fig:resid2}. The top panel displays residuals in the in-band energy range ($R_{BGnet}$), and the bottom panel shows residuals at 0.3--0.4 keV ($S1_{net}$).  The night/day observations are distinguished with a black cross / red triangle, respectively.  What can these residual rates tell us about systematic uncertainty when applying the 3C50 background model? Considering first the in-band residuals (top panel), the 3C50 model is shown to be most effective when the background count rates is low. The need to bifurcate the evaluation into high and low count rates is tied to the pattern of points in the top panel.  When $R_{BG} < 2$ c/s, which corresponds to 82\% of all background GTIs, then $R_{BGnet}$ has $rms$ value 0.33 c/s. In Sections \ref{sec:practical} and \ref{sec:lc} below, we show that the quality of these results can be improved by quality-filtering the background-subtracted spectra in off-target energy bands. In Fig.~\ref{fig:resid2}, the 3C50 model residuals above 2 c/s becomes more random, losing the population near zero, in marked contrast with the bottom panel. We interpret this as evidence that an additional model parameter, which has not been identified, has first-order significance when the background rate is high.  Below, we also provide methods to identify and filter out some of the GTIs associated with high background rates, to protect the integrity of \texttt{NICER} science while further studies of the \texttt{NICER} background go forward.

The bottom panel of Fig.~\ref{fig:resid2} shows very small residuals in $S1_{net}$ during \texttt{ISS} nighttime (black crosses), but the daytime GTIs have higher residuals, particularly when the $nz$ rates are high. High residuals in $S1_{net}$ are matched with much higher residuals in $S0_{net}$, and this provides a method to use $S0_{net}$ to safeguard $S1_{net}$ or $A_{net}$, taking guidance for the relationship between noise leaks in $S0, S1,$ and the $A$-band given in Table~\ref{tab:stage2}. Users can screen for problematic GTIs by choosing a maximum tolerable light leak in $S1_{net}$ or $A_{net}$ for their science, finding that level in Table~\ref{tab:stage2} : $S1$ or $A$, and then use the corresponding $S0$ value as the filter criterion for $S0_{net}$ to exclude GTIs that are likely to exceed the chosen noise limit. Systematic differences within any $nz$ bin in the Stage 2 library will leave residuals of either sign in $S0_{net}$, $S1_{net}$, and $A_{net}$, and filtering efforts should mirror the rejection criteria accordingly.

\begin{figure}[ht!]
\includegraphics[angle=-90.,width=5.0in]{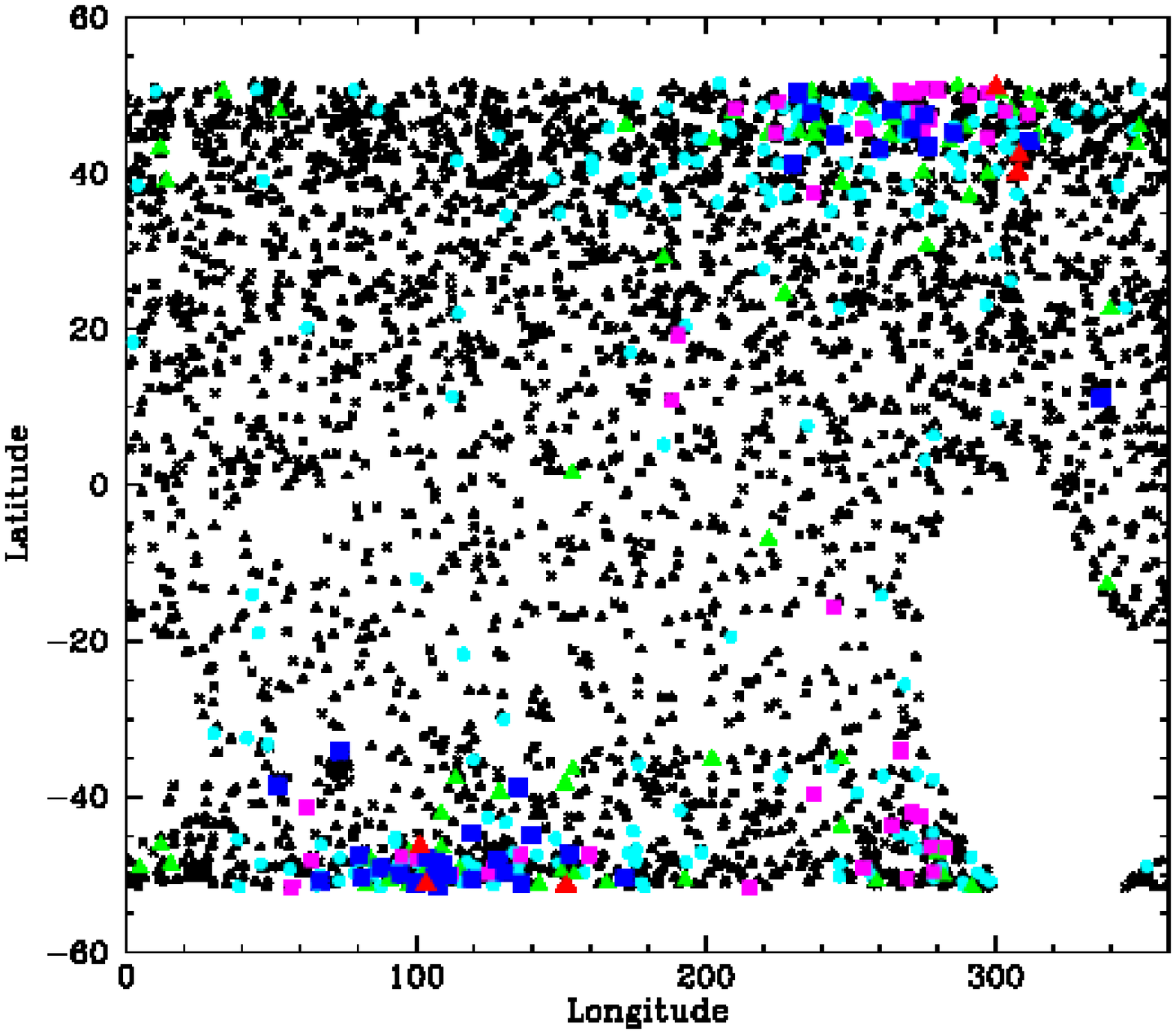}
\caption{Map of the background observations in Earth longitude and latitude, with the magnitude of 3C50 model residuals, $R_{BGnet}$, represented in the symbols type. The c/s ranges and symbols are: $\pm 0.1$ (black ``x''); $\pm 1.0$, excluding the previous level (small black triangle); $\pm 2.0$, excluding previous levels (cyan filled circles); $-5.0$ to $-2.0$ (medium size magenta square);  2.0 to 5.0 (medium green triangle); $< -5.0$ (large red triangle); and $ > 5.0$ (large blue square). The points terminate in latitude at the inclination of the \texttt{ISS} orbit ($52 \deg$). The highest residual count rates from the 3C50 model reside in the polar horns in the \texttt{ISS} orbit.  The region void of points in the lower-right quadrant is the exclusion zone for the SAA, and there are no indications of problems near the chosen SAA boundaries.
\label{fig:residmap}}
\end{figure}

The origin of GTIs with high background model residuals is revealed in Fig.~\ref{fig:residmap}, where different intervals in $R_{BGnet}$ (see the Fig. caption) are plotted on a grid of Earth longitude and latitude, using the orbit location at the midpoint of each GTI.  It is clear that the GTIs with the largest residuals coincide with the polar horns in the \texttt{ISS} orbit. The first two intervals (i.e., residuals within $\pm 1.0$ c/s) account for 90\% of all GTIs, while the intervals with largest residuals (1.2\% of the total and $ |R_{BGnet}| > 5.0$ c/s) are largely (86\%) confined to high latitude: $|lat| > 42.0 \deg$. On the other hand, the \texttt{NICER} exclusion zone for the SAA (Southern area void of points) appears to effectively exclude any background-related problems. This exclusion zone coincides with keyword \texttt{NICER\_SAA} = 1 in the pipeline's information files.

\section{Practical Considerations for the 3C50 Background Model} \label{sec:practical}

\subsection{Implementing the 3C50 Background Model}

The prototype ftool task ''\texttt{nibackgen3C50}'' is the recommended tool to run the 3C50 background model. Although not yet formally part of the HEASoft \texttt{NICER} software suite, it is compliant with NASA HEASARC standards and is supported by the \texttt{NICER} Guest Observer Facility. 
There are two time intervals that are important in the use of \texttt{nibackgen3C50}.  The first is the particular investigation interval for which a predicted background spectrum is desired.  This is driven by the input given to the tool, which is nominally an ObsID,  i.e., the reference number and top level directory for a daily accumulation of GTIs for a single target.  Users can obtain background spectra on timescales shorter than an ObsID (which generally contains multiple GTIs) by alternatively inputting the combination of a single unfiltered event file and GTI file. In either case, one background spectrum is output, per call to \texttt{nibackgen3C50}. Thus, users who plan to investigate target spectra on timescales shorter than one day would sort the unfiltered event files (i.e., the pipeline's \$ObsID/xti/event\_cl/ni*ufa.evt.gz files) into a series of smaller files with the intended time boundaries and the run \texttt{nibackgen3C50} sequentially on these files along with their associated GTI files. 

The second important time interval is internal to \texttt{nibackgen3C50}, which generally computes the background in sub-intervals and then provides the exposure-weighted results in the output file.  The sub-intervals may be the GTIs within the input file, or a shorter timescale directed by the user. Sub-intervals are never allowed to cross GTI boundaries, since the gaps between GTIs can be many hours, with a likelihood of different background conditions on either side of the gap. The background rate systematically varies with the \texttt{ISS} location in its 93 min orbit. As noted above, \texttt{NICER} GTIs are seldom as long 2 ks, and a typical monitoring program has an average GTI $\sim$600 s, or 11\% of the orbit.  As a fraction of the \texttt{ISS} orbit, GTIs are an acceptable choice for \texttt{nibackgen3C50} sub-intervals.  Shorter intervals are also acceptable, but below $\sim$100 s, Poisson noise in $ibg$, the parameter that normalizes the nighttime library selections, can be a concern.

Another option of \texttt{nibackgen3C50} is the ability to control the FPM selections by listing the ones to ignore.  Then, for each sub-interval, \texttt{nibackgen3C50} reads the event lists and calculates the average count rates in $ibg$, $hrej$, and $nz$ for the selected FPMs, scales these rates to the level of 52 FPMs to make library selections (Tables \ref{tab:stage1} and \ref{tab:stage2}), normalizes the nighttime library selection by $ibg / ing_{lib}$ (Table \ref{tab:stage1}), and then, if $nz_{52} > 200$, adds the selected daytime library spectrum, normalized by $nz / nz_{lib}$ (Table  \ref{tab:stage2}). Then, as noted abve, the sub-interval spectra are exposure weighted to produce the modeled background spectrum.  Further information about the control of time intervals and other command parameters for \texttt{nibackgen3C50} are available via the HEASARC \texttt{NICER} tools website \footnote{\url{https://heasarc.gsfc.nasa.gov/docs/nicer/tools/nicer\_bkg\_est\_tools.html}}, from where it may be downloaded and locally installed.

Additional implementation notes are as follows. The 3C50 model enforces a minimum $ibg_{52}$ value of 0.016 c/s, to avoid cases where short intervals and low $ibg$ rates would lead to occasions when $ibg_i$ is zero, and the background prediction would also be zero, since $ibg_i / ibg_{lib}$ is the  normalization factor for the stage 1 contribution to the predicted background. This lower limit on $ibg_{52}$ was estimated from the distributions in $R_{BG}$ and $ibg_{52}$ for the 3556 GTIs examined in this study. Finally, the current version of the \texttt{nibackgen3C50} tool described above allows for application of the $hbg_{net}$ quality check, with the user selecting the maximum allowed absolute value. The $SO_{net}$< check is not implemented the current release, but will be in subsequent releases.

The \texttt{NICER} data archive contains event files that were calibrated with a series gain solutions, which can be identified by the keyword ''GCALFILE''. Users are recommended to bring their data sets to a uniform calibration level with the ''nicerl2'' tool, using either the calibration used in this paper, nixtiflightpi20170601v005.fits (2020), or the more recent one nixtiflightpi20170601v006.fits (2021), for which there are no differences below 12 keV. The \texttt{nibackgen3C50} tool attempts to match the GCALFILE in the input event lists with an appropriate set of library spectra, and 3C50 Model libraries have been prepared for two previous gain calibrations, governed by GCALFILEs nixtiflightpi20170601v002.fits (2018) and nixtiflightpi20170601v004.fits (2019)).  By default, if a match cannot be made, then the user is warned, but a background spectrum is produced using the most recent model library.

Tables \ref{tab:stage1} and \ref{tab:stage2} the details in this Section provide sufficient information (i.e., cell boundaries, library normalizations, minimum $ibg_{52}$ value, and GCALFILE issues) for users who wish to download the model library files, which are a simple directory of standard ''pha'' files, and script their own implementation of the 3C50 background model. The calculation steps would be the same as those given above in the functional outline of \texttt{nibackgen3C50}, and the scripts would need to include commands to sort out the events corresponding to definitions of $ibg$, $hrej$, and $nz$, e.g., using ''fselect'' on the unfiltered and calibrated event lists, given choices of time intervals and lists of FPMs (''DET\_ID'') to exclude.  The conversion of the model-parameter event lists to event rates would complete the assembly of required 3Crp parameters: $N_{FPM}$, $ibg$, $hbg$, and $nz$, for each time interval needing a background prediction.

\subsection{Variations in {\it ibg}, {\it hrej}, and {\it nz} within a GTI}

The selection of GTIs to populate the model libraries did not have an explicit screening step for variations in the values of model parameter within a GTI (Table~\ref{tab:select}). We address this issue here. To evaluate the 3C50 parameter values for each GTI interval, we extracted both the spectra and light curves (1 s bins) for each parameter.  The light curves were routinely used to calculate the mean ($\mu$), standard deviation ($\sigma$), and the variance in excess of Poisson statistics ($\sigma_{int}^2  = \sigma^2 - \mu$) for each parameter.  Trends with brightness were investigated, and we explored several filtering criteria and their ramifications.  

For $ibg$ and $hrej$, the low values of $\mu$ make it inappropriate to use the fraction of intrinsic deviations (i.e., sqrt($\sigma_{int}^2) / \mu$) as a screening tool.  Instead, an ad hoc relationship was favored, with a rejection criterion: $\sigma_{int} > 1.0 + 0.2 \times \mu$. At the lowest count rates (see Fig.~\ref{fig:ampl}), the intrinsic deviations must be above 1 c/s to prompt rejection, while at the highest rates (i.e., 10 c/s), the intrinsic deviation must be 20\% of the mean rate.  Considering all of the GTIs with average parameter values within the 3C50 model limits, $hrej$ variations fail this test in only 0.8\% of the intervals, while the $ibg$ rejection rate is 5.4\%, and this group includes all of the $hrej$ failures. Furthermore, all but 50 of the $ibg$ failures were already rejected for library use by other criteria listed in Table~\ref{tab:select}. These remaining 50 cases are all among the 183 GTIs in the brightest three levels of the night library, i.e., in cells 51-75. Rather than exclude these cases, we concluded that variability in $ibg$ is another characteristic of the high background conditions that do not fall in line with the 3C50 model.

Variations in $nz$ are an entirely different matter. The count rates are high, and systematic differences in the response of individual FPMs to the optical light leak may occur during \texttt{ISS} daytime. The effects of the light leak on the X-ray spectrum are not the same, at a given FPM-integrated count rate in $nz$, if the distribution is skewed toward one FPM rather than being more evenly distributed.  The most striking example is FPM \#34, which was eliminated from this study because of its frequent extreme response to \texttt{ISS} daytime. In practice, it was found that the disparity in FPM noise rates was a more important issue than the changes in $nz$ within a given GTI. Different rejection criteria were investigated, using 3C50 model residuals in the S1 and A bands as the metrics for quality assessment. We adopted the criterion that a GTI would be excluded from consideration (during \texttt{ISS} daytime) if $FPM_i$ with the highest noise rate yields a GTI-average rate $nz_i > 100$, while $FPM_i$ also contributes more than 15\% of the total noise counts. This step is included in the data selection outline given in Table~\ref{tab:select}. GTIs that fail this test were rejected from further consideration , in order to maintain the strategy to use the same 50 FPMs to build the model libraries.  For general investigations with \texttt{NICER}, users could alternatively choose to excluding the offending $FPM_i$ and recompute  the extractions from the event lists for that GTI with a reduced set of selected FPMs.

\subsection{Filtering Steps After Background Subtraction to Improve Data Quality}

To deal with systematic errors in the background subtraction process, the strategy was introduced in Section~\ref{sec:eval} to filter out results on the basis of residual count rates in spectral bands that are not needed for science.  The energy range for any spectra extracted from cleaned event lists is 0.2--15.0 keV, and in the context of this background investigation, this can be seen as $S0 + S1 + A + B + C + D + gap + hbg$, corresponding to energy bands 0.2--0.3 0.3--0.4, 0.4--1.0, 1--2, 2--4, 4--12, a gap, and 13--15 keV.  Background-subtracted rates are expected to be near zero in $S0_{net}$ and $hbg_{net}$, while the other bands contain the target spectrum to be used for science analyses.  Tabulating the count rates in $S0_{net}$ and $hbg_{net}$ provides a basis for quality fitering the background modeling process.  One can view $hbg_{net}$ as a quality diagnostic for the Stage 1 background component, while $S0_{net}$ is a diagnostic for the Stage 2 component. 

Three levels of filtering are advised for \texttt{NICER} investigations of targets with different levels of X-ray brightness.  They are detailed below in the sense of data selections for quality purposes, to be applied to GTIs prior to science analyses of light curves or spectra. The next Section offers two examples of this process. The filter levels given below are illustrative, and users should explore the tradeoffs in coverage versus data quality to decide the optimal filter criteria that are consistent with the investigation goals and requirements.

\begin{itemize}
    \item Level 1 filter selects GTIs with (($-30.0 < S0_{net} < 30.0)$ c/s \&\& $(-0.5 < hbg_{net} < 0.5$)) c/s.  This filter should be applied to even the brightest X-ray sources.
    
    \item Level 2 filter selects GTIs with (($-10.0 < S0_{net} < 10.0)$ c/s \&\& $(-0.1 < hbg_{net} < 0.1$)) c/s. The level 2 filter is appropriate for moderately bright sources, e.g., $20.0 < R_{net} < 300$ c/s. For moderately bright sources with very soft spectra (e.g., with detections limited to energy below 2 keV), filter level 2S can additionally impose: $-0.5 < D_{net} < 0.5$ c/s, where the D-band (4--12 keV) is given up as an additional background band.
    
    \item Level 3 filter selects GTIs with (($-2.0 < S0_{net} < 2.0)$ c/s \&\& $(-0.05 < hbg_{net} < 0.05$)) c/s. This filter is appropriate for faint sources, e.g., $R_{net} < 20.0$ c/s). For a faint source with a very soft spectrum, filter level 3S can again impose: $-0.5 < D_{net} < 0.5$ c/s.
    
    \item It has been shown that the majority of GTIs with the highest background rates and largest model residuals occur in the polar regions of the \texttt{ISS} orbit (Figs.~\ref{fig:resid2} and \ref{fig:residmap}). However, there is no effective way to screen results by orbit position without incurring significant data losses. To illustrate this, we define the a group of ``bad'' model residuals as 153 GTIs (of 3477) with $R_{BGnet} < -2.0$ or $R_{BGnet} > 2.0$ c/s.  A broad definition of the polar region, with latitude $lat < -42 \deg$ or $lat > 42 \deg$, captures 80\% of the bad GTIs.  However, only 9.4 \% of the polar GTIs are bad, and GTI exclusion on this basis would be costly.  An ad hoc definition of the polar horns can be made from Fig.~\ref{fig:residmap}, using the same polar region with additional constraints for longitude intervals: $200 < lon < 320$ (North) and $60 < lon < 180$ (South).  This region captures 63\% of the bad GTIs, but only 19 \% of the GTIs are bad.
    
\end{itemize}

\subsection{Model and Filtering Considerations for the Brightest Source}

Considerable attention has been paid to the brightest and the softest X-ray sources observed with \texttt{NICER}, to investigate the effects of source counts on the background parameters and also the background-subtracted count rates pertinent to quality filtering.  We first consider the the high-energy range of the spectrum, specifically $ibg$, a 3C50 model parameter (15-18 keV, in-focus), and $hbg_{net}$ (13-15 keV, in-focus), which used as a data quality filter. We note that $ibg$ is a raw measurement, while $hbg_{net}$ is a background-subtracted quantity (subscript "net"). Both of these energy bands are outside the imaging effective areas of the concentrator, but there is still a finite probability that a high energy photon may pass straight through to the detector, without interacting with the concentrator foils.  Thus, it is relevant to investigate whether any extremely bright X-ray sources may elevate the count rates of either parameter.  Of particular interest are Scorpius X-1, the brightest X-ray source in the sky (116,000 c/s when normalized to 50 FPMs), the black hole transients, MAXI J1820+070 (65,000 normalized c/s at maximum) and MAXI J1348-630 (47,000 normalized c/s), and the neutron star transients with high-mass companion stars (HMXBs) and relatively hard X-ray spectra, Swift J0243.6+6124 (28,000 c/s at maximum) and A0535+26 (6,000 c/s). The first three cases are the only targets (2017-2020) for which there was a commanded reduction of the number of active FPMs, so that the telemetry rate would remain below the maximum event rate ($\sim$ 30,000 c/s) for the cables connecting the output of the MPUs to the telemetry stream. For all of these sources, we find that contamination of $ibg$ is not an issue. Values of $ibg$ are found to be uncorrelated with changes in source intensity, when comparing these quantities on the GTI timescales. Variations in $ibg$ are dominated by seemingly random changes in the background conditions, with no evidence of the outburst profile of the X-ray sources.

However, the impact of these bright or hard sources on $hbg_{net}$ is somewhat different.  After background subtraction, a residual count rate 
in $hbg_{net}$ is seen in Sco X-1 (up to 1 c/s), and similar residuals are seen for the pair of bright HMXBs at times of maximum intensity. All of the other bright transients show $hbg_{net} < 0.2$ c/s when the sources are near maximum intensity.  Since the prescription for the level 1 quality filter is to reject background-subtracted GTIs with $hbg_{net} < -0.5$ or $hbg_{net} > 0.5$, observations would be falsely rejected, at level 1 filtering, for the Sco X-1, Swift J0243, and A0535-26. The solution to this problem is to either refrain from filtering these three sources, near times of maximum intensity, or to predict the \texttt{NICER} background with the ''Space Weather'' Model (see Section 9.3), which has no parameters related to measured count rates.  Users of the 3C50 background model are advised to compare the light curves for exceptionally bright or hard X-ray sources with the light curve of derived $hbg_{net}$ values in order to determine customized filtering values that are appropriate.  Furthermore, the level of filtering should be approached as a function of source count rate, particularly for transients that \texttt{NICER} observes with more than 5 magnitudes of dynamic range between intensity maxima and the final measurements as the source returns to quiescence. The increased susceptibility to X-rays from bright and hard sources for $hbg$ (13--15 keV), relative to $ibg$ (15--18 keV), can be understood as a combination of the decreasing absorption cross section at 13--18 keV in silicon, combined with the decreasing photon spectrum in that same range, for most X-ray subclasses. 

An analogous search was made for target contributions to $nz$, the raw count at 0.0--0.25 keV, and the filtering parameter, $S0_{net}$, the count rate at 0.2--0.3 keV in the background subtracted spectrum.  The investigation included the same bright sources noted above, plus very soft X-ray transients, e.g., MAXI J0637-430 (6,000 c/s at maximum) and the coronal flares in HR1099 (reaching 675 c/s). It was found that the model parameter $nz$ is dominated by variations in the Sun angle and is not significantly affected by exceptionally bright or soft source.  However, the level 1 filtering condition, $S0_{net} < 30 c/s$ is exceeded in the brightest GTIs for four sources: MAXI J1820, Sco X-1, MAXI J0637, and HR1099. Again, users can suspend data filtering near times of maximum intensity for these sources, or alternatively they can use the "Space Weather" background model (Section 9.3). The comparison of light curves in $S0_{net}$ versus the broadband source intensity (0.4 -- 12 keV) is a prudent step in the effort to customize filtering and optimize data quality for exceptional sources. \\

\section{Background-Subtracted Light Curves} \label{sec:lc}

\subsection{Observations of the Crab Nebula}

The Crab Nebula is commonly used as a bright reference source in X-ray astronomy.  However, the Crab intensity is not truly constant; long term variations up to 7\% were detected in multi-satellite observations \citep{wils11}.  \texttt{NICER} observations (with target name ``PSR\_B0531+21'') over the interval 2017 August 5 to 2020 April 27 were reprocessed, applying the same calibrations used for the background fields (Section~\ref{sec:data}).  The query for GTIs, again using \texttt{nimaketime} while excluding the undershoot/overshoot rate filters, netted 418 GTIs with duration $> 50$ s. 

\begin{figure}[ht!]
\includegraphics[width=4.5in]{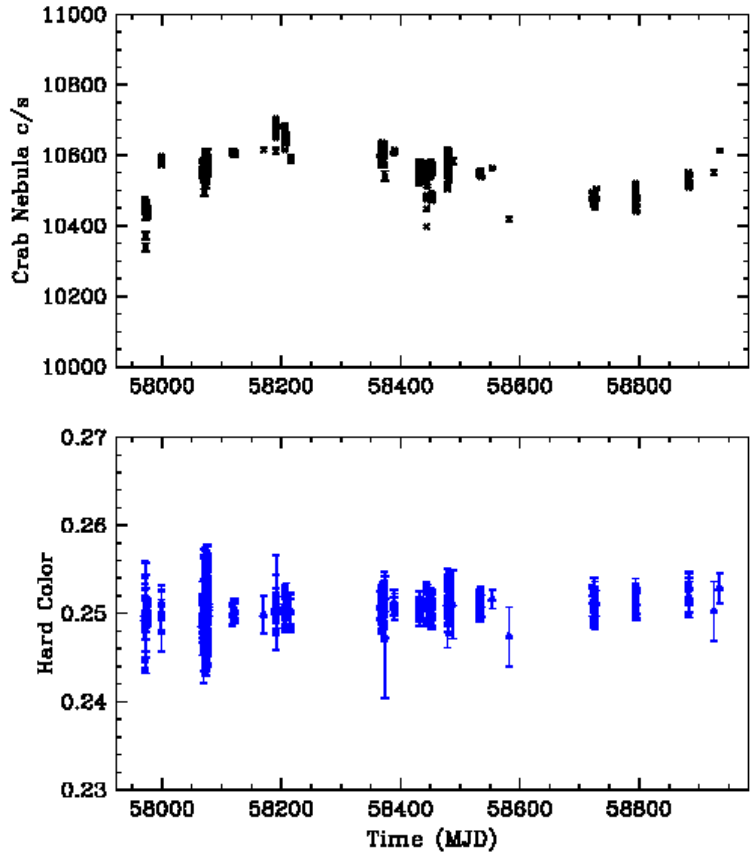}
\caption{top panel: \texttt{NICER} light curve of the Crab Nebula at 0.3--12 keV, after background subtraction with the 3C50 model. The plot shows 408 GTIs with an average exposure of 816 s.  The selected sample has a mean: $10531 \pm 60$ c/s, and most of the the 0.6\% variations can be seen as a gentle 900 day wave in the light curve. Observations with other satellites are needed to determine whether these variations in the Crab are real. bottom panel: the Hard Color, defined as the ratio of count rate at 4--12 keV, relative to that at 2--4 keV. Measurements have a mean and rms, $0.251 \pm 0.001$.
\label{fig:crab}}
\end{figure}

Application of the 3C50 model yielded background predictions for 416 GTIs, while the remaining two cases had $ibg_{52}$ values that exceeded the model limits (10 c/s). Count rates were integrated from the background-subtracted spectra in the range of 0.3--12 keV, and Level 1 filtering was applied (see Section~\ref{sec:practical}).  This filtering step excluded eight GTIs as quality risks, and the count rates for the remaining 408 are shown in Fig.~\ref{fig:crab}, scaled to 50 FPMs.  These data display mean and rms values: $10531 \pm 60$ c/s. In contrast, the eight filter-eliminated GTIs average $10414 \pm 226$ c/s at 0.3--12 keV.  Fig.~\ref{fig:crab} also shows the hard color, which is the ratio of count rates at 4--12 and 2--4 keV (or D/C in terms of energy band labels).  The hard color measurements have a mean and rms, $0.251 \pm 0.001$.  The hard color results indicate the photometric precision that can be achieved with \texttt{NICER} spectra, using modest quality filtering, over the 2017-2020 time interval.

The Crab light curve shows that the 0.6\% variations have a systematic temporal profile that can be seen as a gentle $\sim 900$ d wave in intensity. These results are not corrected for deadtime, but the latter depends on the total event rate (all energies and all event flags), and the total event rate for the Crab varies on an annual timescale, due to the correlation between the noise rate and the solar angle. Observations of the Crab by other space missions will help to determine whether the changes in the \texttt{NICER} light curve are intrinsic to the Crab or arise from systematic factors that have escaped the current investigation. 

\subsection{Light Curve of 1E 0102.2–7219}

The supernova remnant in the Small Magellanic Cloud (SMC), 1E0102.2$-$7219 (hereafter ``E0102''), is a faint calibration source that serves as a flux and spectral line reference for many X-ray instruments \citep{pluc17}. The \texttt{NICER} observations from 2017 July 17 to 2020 June 12 netted 965 GTIs with an average exposure of 438 s. The 3C50 model yielded 941 background subtracted spectra, while filtering steps (see Section~\ref{sec:practical}) left 916 GTIs at level 1 and 804 GTIs at level 2.  We note that the fraction removed by the level 2 filter (15\%) is larger than normal, because E0102 is observed in a wider range of conditions, as a calibration source, compared to many \texttt{NICER} science targets. 

\begin{figure}[ht!]
\includegraphics[angle=-90.,width=5in]{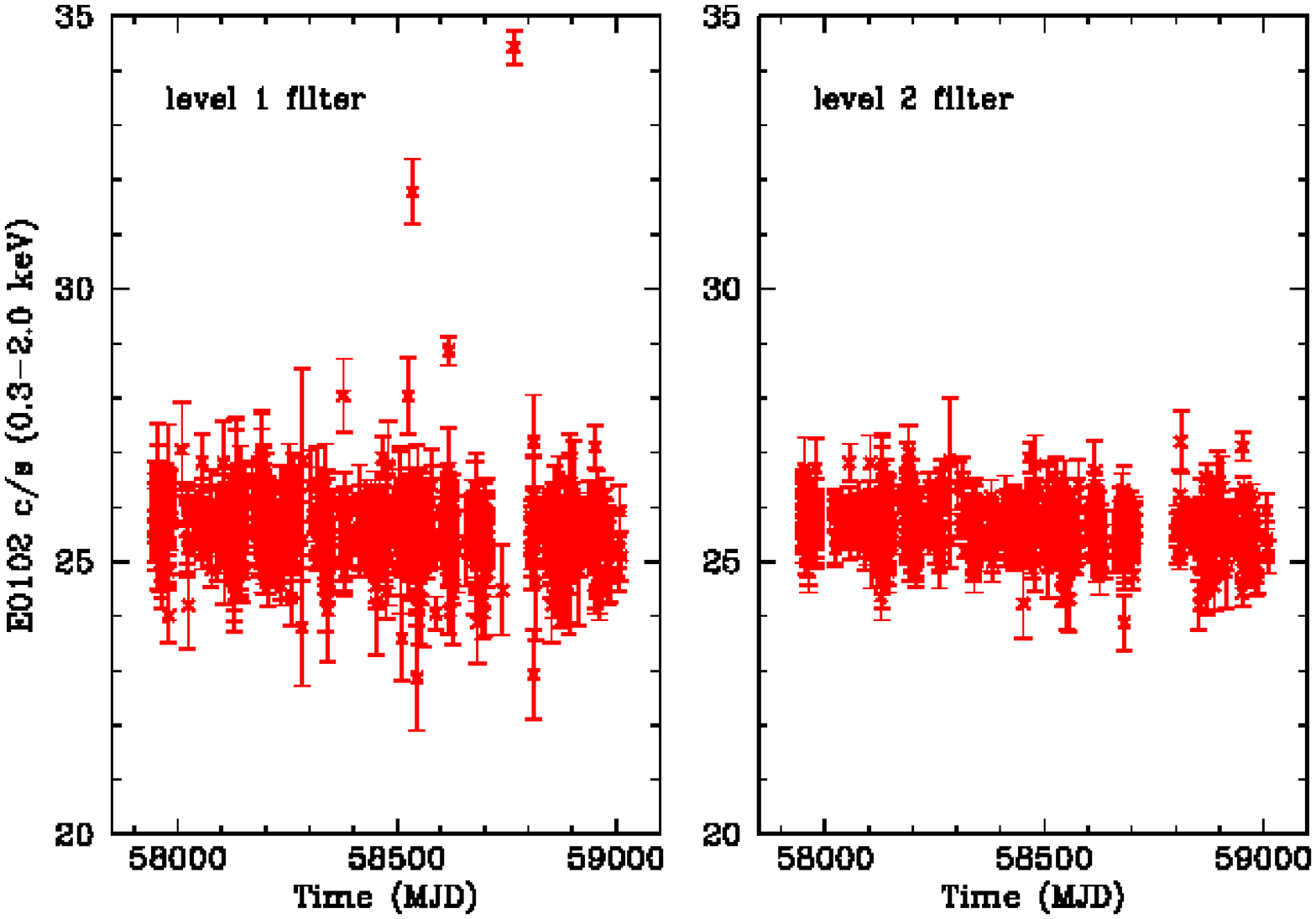}
\caption{\texttt{NICER} observations of the SNR in the SMC, E0102, yielded background subtracted spectra with the 3C50 model for 941 GTIs through 2020 June 12, accumulating 408 ks exposure time.  The soft light curve (0.3--2.0 keV) extracted from all of the GTIs is shown in the left panel, while the same data with level 2 filtering (804 GTIs, 352 ks) is shown in the right panel (see Section~\ref{sec:practical}).  These measurements have mean and rms: $25.60 \pm 0.62$ c/s and $25.63 \pm 0.42$ c/s, respectively, while the rms statistical uncertainty is 0.32 c/s.
\label{fig:e0102}}
\end{figure}

Since the source is soft, we examine the background-subtracted light curve in the range 0.3--2.0 keV.  Fig.~\ref{fig:e0102} shows the results for level 1 filtering (left panel) and level 2 filtering (right panel).  The measurements have mean and rms: $25.61 \pm 1.15$ c/s and $25.62 \pm 0.41$ c/s, respectively. The average statistical uncertainty at 0.3--2.0 keV is 0.3 c/s.  This demonstrates the utility in using $S0_{net}$ and $hbg_{net}$ as metrics for filtering GTIs, sacrificing some amount of temporal coverage to improve quality.

\section{Background Modeling at 1 s Timescale} \label{sec:fast}

The background parameters, $ibg$ and $hrej$, normally have count rates below 1 Hz, and one must integrate for a few hundred seconds to produce an average value with reasonable statistical precision.  However, the occasional surges in the background rates show corresponding variations in $ibg$ and $hrej$ and the relationship between these quantities can help to diagnose whether rapid changes in \texttt{NICER} light curves may originate from either the X-ray target or the in-band background.

\begin{figure}[ht!]
\includegraphics[width=3.5in]{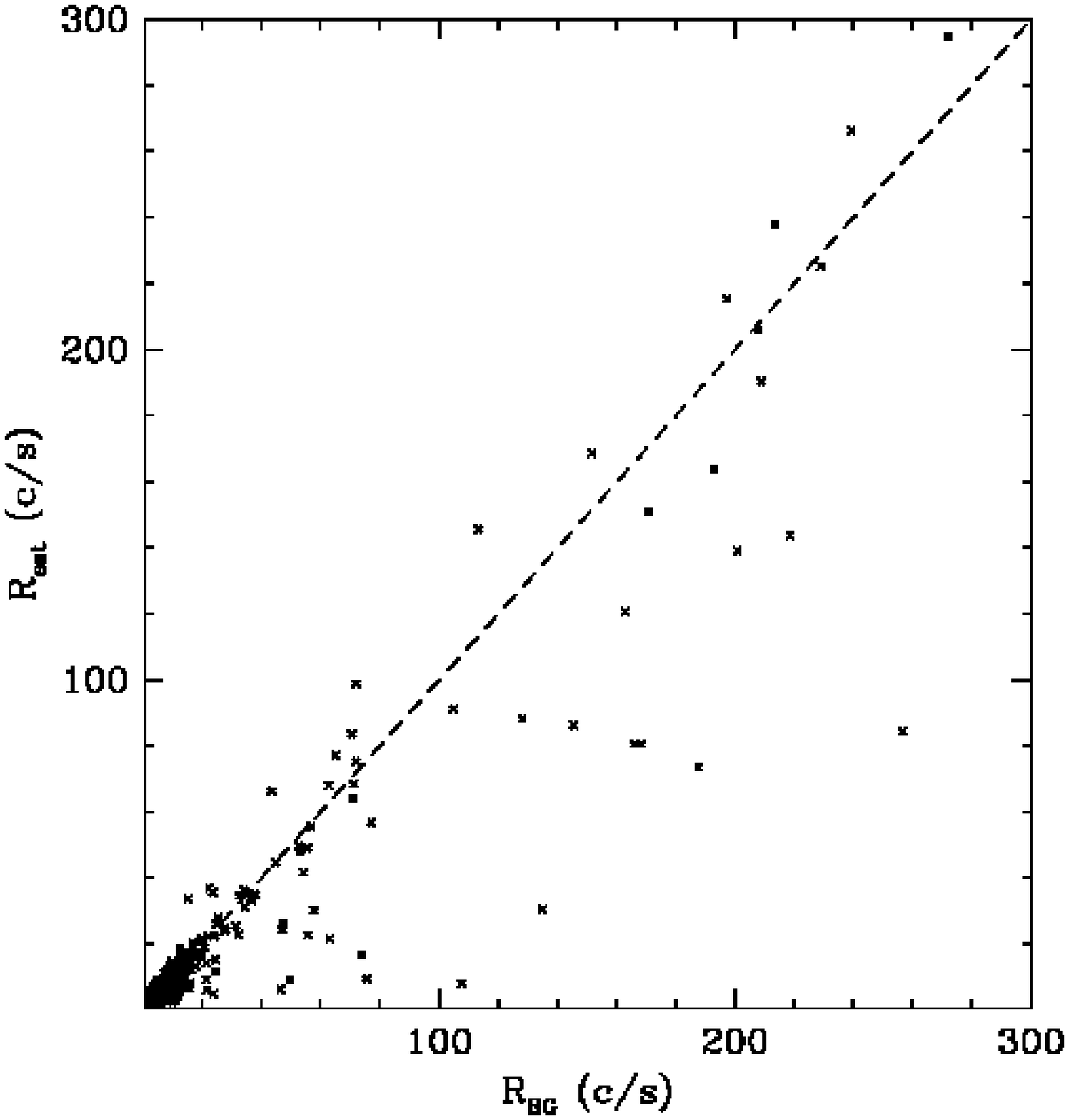}
\caption{Estimate of the in-band background rate using the relationship, $R_{est} = 2.91 * ibg + 4.67 * hrej$. Each point corresponds to one GTI, and the equation is determined with a least-squares fit confined to the range $0.5 < R_{BG} < 300$ c/s.  The dashed line is a reference for the match between this relationship and the measured rate ($R_{BG}$).  The estimator is more precise at low count rate, showing again the systematic error for the 3C50 model when the background rate is high.  Nevertheless, the estimator is an effective aid to diagnose whether observed variability at short timescales originates from the target or from activity in the background.
\label{fig:bg1s}}
\end{figure}

In Fig.~\ref{fig:ampl} it was shown that $ibg$ and $hrej$ are both roughly correlated with $R_{BG}$, with somewhat different average slopes. This motivates a strategy to estimate $R_{BG}$ as a linear combination of $ibg$ and $hrej$ with different coefficients. Using a least-squares fit confined to the range $0.5 < R_{BG} < 300$ c/s, the best fit relationship is $R_{est} = 2.91 * ibg + 4.67 * hrej$, with results shown in Fog.~\ref{fig:bg1s}.  There is significant scatter in the ability of the background estimator to predict $R_{BG}$ at high count rates, pointing to the same problem seen with the 3C50 model residuals (Fig.~\ref{fig:resid2}).  Nevertheless, the background estimator might show rapid increases and temporal structure that resemble the \texttt{NICER} light curve in short time bins, and this would convincingly implicate the background as the origin of the fast flares.

\begin{figure}[ht!]
\includegraphics[angle=-90.,width=5in]{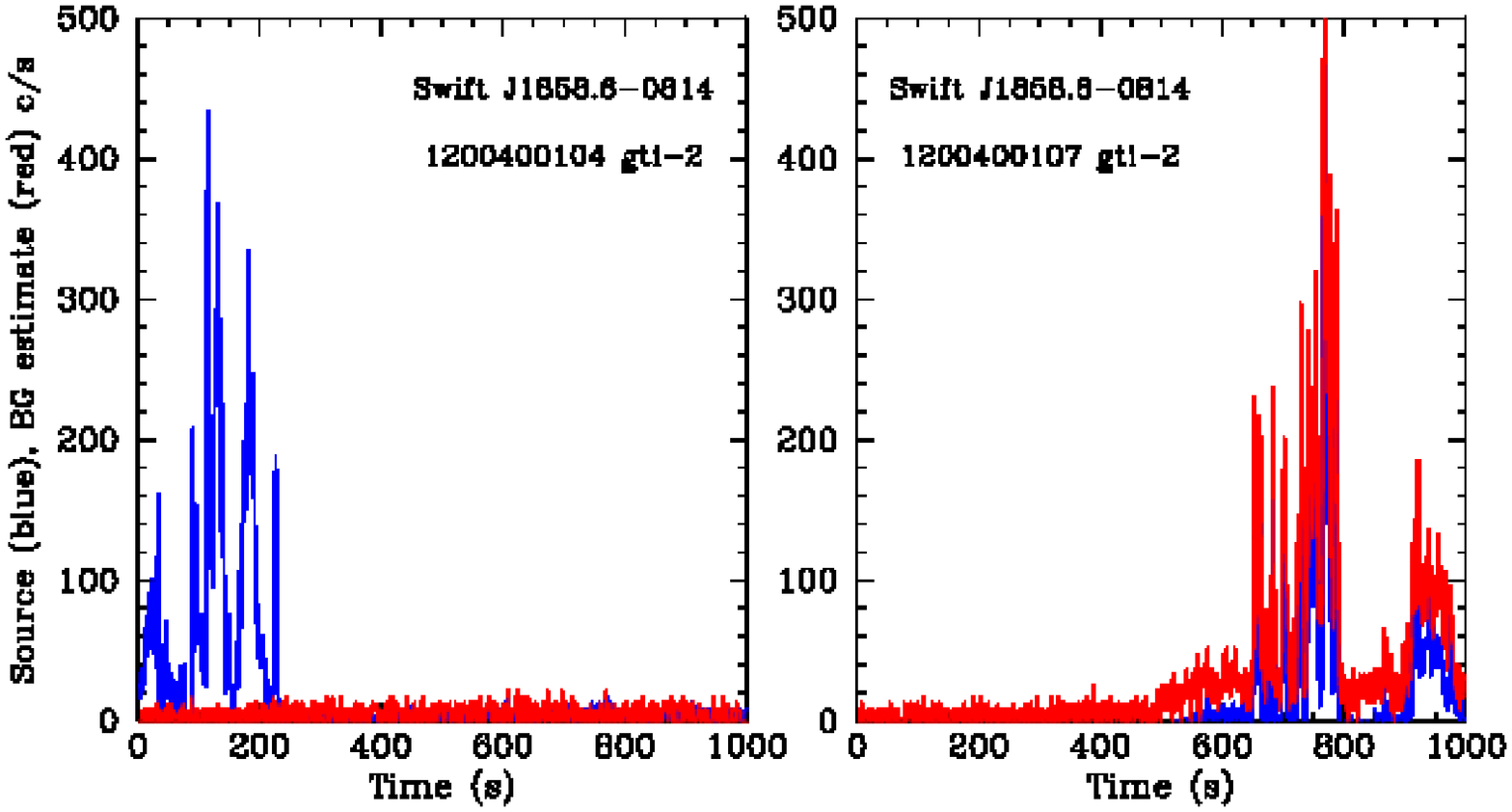}
\caption{Application of the 1-s background estimator to two GTIs from observations of SwiftJ1858 when the source, an X-ray binary system containing an accreting neutron star,  was in a flaring state.  The first GTI shows that the rapid flaring is from the X-ray source (blue), and this behavior has been observed many times with \texttt{NICER} and other instruments.  However, the second GTI shows a light curve with unusual rapid flaring in the background (red), demonstrating that this particular sequence of flares is from the background, rather than the X-ray source.
\label{fig:bgswiftj1858}}
\end{figure}

To illustrate the use of the background estimator, we consider the case of the X-ray transient source, Swift J1858.6$-$0814 \citep{krim18}, hereafter ``SwiftJ1858''. \texttt{NICER} observations from 2018 November 1 through 2019 November 17 show dozens of GTIs containing fast variability, when the source intensity ranges  from non detectable levels to multiple sharp maxima in the range 100 to 1600 c/s \citep{ludl18}. It was later shown that much of this "flaring" is actually driven by variable absorption along the line of sight to a nearly eclipsing binary system \citep{buis21}. After a data gap imposed by low Sun angle during the interval MJD $58805 - 58903$, \texttt{NICER} found SwiftJ1858 to be in a more conventional state of quasi-steady emission with eclipses and absorption dips. Then type I X-ray bursts were detected, identifying the source as an accreting neutron star \citep{buis20}.

Despite the propensity of Swift J1858 to vary rapidly, during the first part of its outburst, it is necessary to distinguish the few cases in which rapid flaring originated in the background, rather than the X-ray source. Fig.~\ref{fig:bgswiftj1858} shows two GTI light curves, in 1 s bins, with contrasting findings regarding the origin of the fast flares.  In both cases, the light curve shown in blue is the background-subtracted count rate using the 3C50 model with parameters averaged over the respective GTIs.  Values for the background estimator are shown in red.  In the first case (the second GTI on MJD 58426), the background intensity remains low and quiet, implying that the flares originate in SwiftJ1858. 

In the second case (the second GTI on MJD 58429), the BG estimator show that the high-amplitude variations coincide with significant activity in $ibg$ and $hrej$.  A precise match is not expected between blue and red curves, since an effort has been made to background-subtract the blue curve on the GTI timescale (note the negative values at times before 800 s), and since the predictability of the background is compromised by systematic error in the model at high background rates.  Nevertheless, it is clear that the second set of flares originated in the background and not in SwiftJ1858. In this example, the background flares did occur in the Southern polar horn, during a time interval with a range in orbit latitude and longitude: $-51.7 \deg < lat < -47.5 \deg$ and $79.0 \deg < lon < 107.2 \deg$.

\section{Discussion} \label{sec:disc}

\subsection{Sensitivity Limits with the 3C50 Background Model}

Quantitative analysis of the residuals in a background model, when applied to observations of blank sky regions, provides information on both the instrument sensitivity limits and systematic problems regarding model performance.  Given the large range in $R_{BG}$ (Fig.~\ref{fig:ampl}) and the limited success of the 3C50 model when $R_{BG} > 2.0$ (Section~\ref{sec:eval}), these topics must be approached with qualifications. After applying the 3C50 model to 3447 GTIs with model parameters within limits, the residuals at 0.4--12 keV are within $\pm 0.5$ c/s in 80\% of the GTIs.  Applying level 3 filtering criteria (see Section~\ref{sec:practical}), which excludes 15\% of these GTIs, the rms value of $R_{BGnet}$ is 0.40 c/s.  This, in turn, implies a detection limit (3 $\sigma$ in a single GTI) of 1.20 c/s at 0.4--12 keV, which is equivalent to $3.6 \times 10^{-12}$ erg cm$^{-2}$ s$^{-1}$ at 0.4--12 keV. assuming the spectral shape of the Crab Nebula. In the soft X-ray band, the corresponding detection limit ($3 \sigma$, single GTI) is 0.51 c/s at 0.3--2.0 keV. These limits would improve by a factor of 4 if the exploratory GTIs for a given target accumulate 10 ks of exposure. We note that the percentage of GTIs that pass level 3 filtering should exceed 90\%, since the scheduling of science targets would be more favorable than the program to widely sample the background conditions.

\subsection{Model Limitations at High Background Rates}

Our assessment of the 3C50 model noted evidence of a missing model parameter that is particularly important at high values of the raw background rate at 0.4--12 keV ($R_{BG}$).  The correlation between $R_{BG}$ and $ibg$  implies that the most of these events are associated with the in-focus component implied by values in \texttt{PI\_ratio}.  This is reminiscent of the early {\it Chandra} discovery that protons could scatter off the mirrors and come into focus on the detectors when the satellite passed through the radiation belt \citep{odel10}. The hypothesis that a similar condition is affecting \texttt{NICER} would imply that the background model should consider the angle between the camera pointing direction and the local magnetic field lines during the course of the \texttt{NICER} orbit.  Such attention was suggested by \cite{fuka09} in the background model for the {\it Suzaku} HXD Instrument.

Further motivation to track the camera viewing direction, relative to its position in the Earth orbit, is provided by a study of an archive of particle rate measurements built with a series of NOAA polar-orbiting satellites \citep{fida10}.  Since 1998, these satellites have been equipped with the Space Environment Monitor (SEM-2) system, which contains two sets of instruments that monitor the energetic charged particle environment above the Earth. One of these systems, the Medium Energy Proton and Electron Detector, has two sets of detectors mounted perpendicular to each other.  The different viewing angles produce differential particle fluxes for electrons in the ranges of both 30-100 and 100-300 keV (see Fig. 1 of \cite{fida10}). The pitch angle, i.e., the angle between the charged particle flow and the local magnetic field, varies systematically with the position in low Earth orbit, suggesting that the particle flux should depend on the longitude, latitude, the camera angle with respect to the local magnetic field, and the local pitch angle.  This context will be explored further in the effort to link the direction of particle flow to the in-focus component in the \texttt{NICER} background.

The limitations of the 3C50 model were first apparent in Fig.~\ref{fig:cellrbg}, where the raw background rates showed significant variations when binned in model cells for the nighttime library.  This, in part, motivated the strategy to re-normalize the selected library spectrum for a given GTI$_i$, by the factor $ibg_i / ing_{lib}$. An assessment of the impact of this step can be made by comparing the residuals with and without the re-normalization step, while making use of the quality filtering criteria described in Section 6.3.  Closing the loop on the background pointings with (without) $ibg$ re-normalization leaves residuals that pass level 1 filtering in 93\% (88\%) of all GTIs, while passing level 2 filtering 85\% (85\%) of the time. The conclusion is that $ibg$ re-normalization does provide a modest improvement in the quality of the 3C50 model when the background rate is high.

\subsection{Comparisons with Parameters of the Environmental Background Model}

The ``Space Weather'' Background Model is an alternative to 3C50 that predicts \texttt{NICER} background spectra on the basis of the local spacecraft environment (Gendreau et a. in prep.), and it is implemented in the FTOOL, ``nicer\_bkg\_estimator'', which is also available via the HEASARC \texttt{NICER} tools website (prior footnote).  The principal model parameters are the local cutoff rigidity (``COR'' ; \cite{smar05}), which is a measure of the shielding provided by the Earth's magnetic field, and the $K_p$ index, which is a global measure of disturbances in the magnetic field.  The \texttt{NICER} pipeline furnishes the values of ``$COR\_SAX$'', every second, in the filter files, and the values of the cutoff rigidity are computed with a particular model developed for the $BeppoSAX$ Mission \citep{amat02}. $COR\_SAX$ shielding has units in the range $0-17$, while the $K_p$ index is given for each 3 hr interval, with a range $0-6$ quantized in steps of 0.333.  Values for the $K_p$ index are obtained from the GFZ site in Potsdam (https://www.gfz-potsdam.de/en/kp-index/).

\begin{figure}[ht!]
\includegraphics[width=5in]{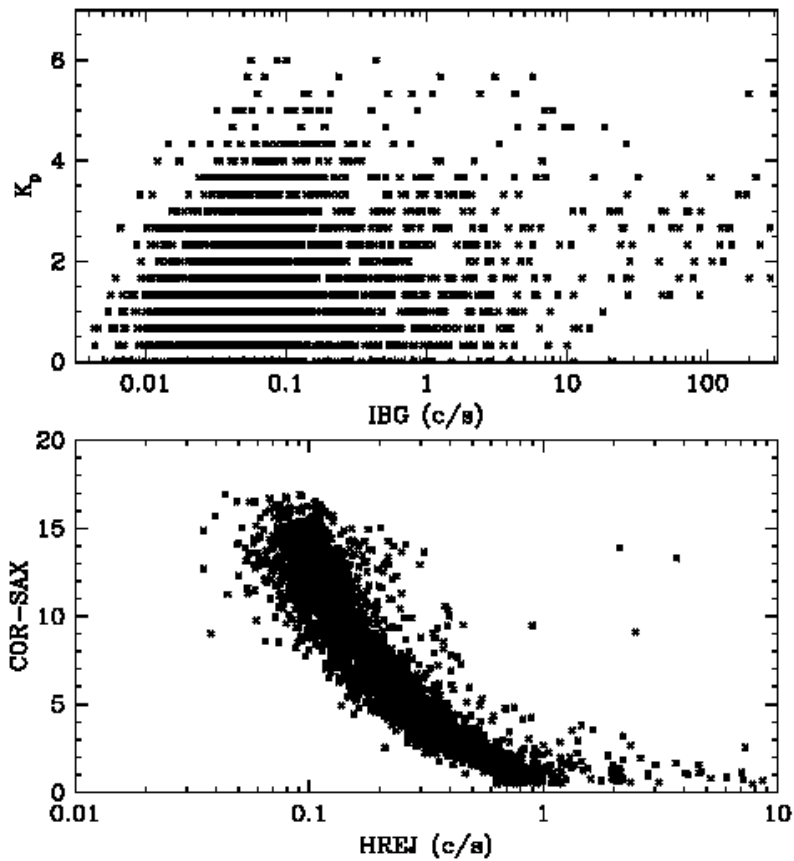}
\caption{Relationship between 3C50 model parameters and those of the \texttt{NICER} Environmental Model.  The plot of $hrej$ vs. the cutoff rigidity ($COR\_SAX$) is the best correlation seen between any two hypothetical background parameters in this study. On the other hand $ibg$ and $K_p$ pair with their model partners in different ways.  
\label{fig:env}}
\end{figure}

The relationship between model pairs for the 3C50 and Space Weather models is examined in Fig.~\ref{fig:env}.  It is immediately apparent that $hrej$ and $COR\_SAX$ are measuring the same phenomenon, i.e., the amount of magnetic shielding at the \texttt{ISS} position is inversely proportional to the rate of spatially extended events due to particles, as measured with $hrej$. The $COR\_SAX$ parameter would be a desirable substitute for $hrej$, in future versions of the 3C50 model, because it is readily accessible in \texttt{NICER} filter files, and it is free from the statistical accuracy limits that confronts $hrej$, due to its low count rate.

\section{Summary}

\texttt{NICER} has a comparatively low background rate, typically $10^{-4}$ times the broad-band count rate of the Crab Nebula, but it is highly variable in both amplitude and spectral shape. The silicon drift detectors are high-throughput, but single-channel devices, and so the background spectrum must be predicted using measurements that are not affected by the targeted X-ray source. The 3C50 model predicts the background spectrum using an empirical approach and three model parameters. Data analyses are based on recurrent observations of seven pointing directions that are void of detectable sources. Spectra from a wide range of observing conditions are sorted by values of the model parameters to build a two-stage library of spectra that are the core of the background model. It is noted that most particle hits are automatically excluded from either the target spectrum or the background model because the energy of the event trips the overshoot flag, removing such events from spectral consideration.

Two model parameters, $ibg$ and $hrej$, track background components associated with particle-induced events. They are distinguished by values of \texttt{PI\_ratio}, which is the ratio of event energies in the fast measuring chain, relative to the slow chain in the instrument electronics. Values of \texttt{PI\_ratio} can discriminate detector ionization locations near the center of the silicon drift detector (i.e., events appearing ''in-focus'', from those near the outer edges of the detector (hence associated with spatially extended events). We define $ibg$ as the rate of in-focus events at 15--18 keV (beyond the effective area of the optics), while $hrej$ is the rate of particle events at 3--18 keV that originate near the outer edges of active silicon, underneath the metal collimator. A grid of values in these two parameters is used to bin and average the GTI-based collection of background spectra to form the stage 1 library of the model. Measured values in these parameters for any given target observation are then used to select a matching library spectrum. That spectrum is re-normalized by $ibg / ibg_{lib}$ to form the stage 1 prediction of the \texttt{NICER} background. The third parameter, $nz$ (count rate at 0--0.25 keV), allows predict of a low-energy excess that is tied to observations conducted in sunlight, when $nz > 200$ c/s. Twelve intervals in $nz$ are used to sort and average the residual spectra from the stage 1 process, applied to all of the background observations, to form the stage 2 library of the model. For target observations, the measured value, $nz_i$ is used to select a spectrum from the stage 2 library.  That spectrum is re-normalized by $nz_i / nz_{lib}$, and the result is added to the stage 1 background spectrum to complete the background prediction. 

The small contribution from the cosmic diffuse X-ray background is carried into the background model by the manner in which the stage 1 library is constructed. This component is always present in the \texttt{NICER} field of view, and its inclusion in the 3C50 model is guaranteed by the imposition of a minimum value for $ibg$, 0.016 c/s. There are no provisions in the model for diffuse Galactic emission, components local to the Earth or the solar system, or contaminating sources in the field of view. Such contributions, when anticipated, must be considered externally.

An examination of 3556 GTIs, with an average duration of 570 s, shows that the in-band cont rate of good events at 0.4--12 keV, scaled to 50 selected detectors, has a median value 0.87 c/s. However, the distribution is quite broad, ranging from 0.33 to 300 c/s, after excluding 1\% outliers on each end. After applying the 3C50 model to 3447 GTIs with model parameters within limits, the residuals at 0.4--12 keV are within $\pm 0.5$ c/s in 80\% of the GTIs.  However, residuals persist at 20--30\% of the initial rate for the brightest cases, which tend to occur in the polar horns of the \texttt{ISS} orbit (mixed with many quiet GTIs at the same polar locations).  The inaccuracy of the model, when the background rate is high, suggests one or more missing model parameters. Quality filtering criteria are developed to warn users when the predicted background spectrum is not likely to be satisfactory.  When such filtering criteria are applied at the level appropriate for faint X-ray sources, the systematic uncertainty in the model, which is an estimate of the detection limit, is 1.20 c/s at 0.4--12 keV (3 $\sigma$, for a single GTI), and 0.51 c/s at 0.3--2.0 keV. For a Crab-like spectrum, the detection limit at 0.4--12 keV  is equivalent to $3.6 \times 10^{-12}$ erg cm$^{-2}$ s$^{-1}$.  The limiting count rate in soft X-rays is equivalent to $4.3 \times 10^{-13}$ erg cm$^{-2}$ s$^{-1}$, assuming a 100 eV blackbody spectrum, with an ISM column density of $5 \times 10^{20} cm^{-2}$.  These limits would improve by a factor of 4 if the exploratory GTIs accumulate 10 ks. The GTIs that pass such filtering criteria amount to 85\% of the total, while higher success rates would be expected for general targets scheduled more favorably than the background observations.

Under normal conditions, the empirical model's background predictions are limited to timescales of minutes or longer, because of Poisson noise in $ibg$ and $hrej$, which often have count rates $< 1$ c/s. However, the crude background estimator, $R_{est} = 2.91 * ibg + 4.67 * hrej$, can be applied on timescales of seconds or less, to help assess whether observations of fast variability originate in either the source or the background. Background flares that are associated with significant variations in the raw, in-band count rate produce momentarily high values of $ibg$ and $hrej$, and the temporal structure in $R_{est}$ will be highly correlated with the in-band variations under scrutiny.

\end{document}